\documentclass[sigconf, screen]{acmart}
\usepackage{graphicx}
\usepackage{textcomp}
\usepackage{array}
\usepackage{epstopdf}
\usepackage{subfigure}
\usepackage{caption}
\usepackage{multirow}
\usepackage{titlesec}
\usepackage{setspace}
\usepackage{algorithm}
\usepackage{algorithmic}
\usepackage{epstopdf}
\usepackage{color}
\usepackage{marvosym}

\usepackage{url}

\newcommand{\ie}{\textit{i.e.,} }
\newcommand{\eg}{\textit{e.g.,} }

\newcommand\opt[1]{}
\newcommand\find[1]{}

\newcommand{\ls}[1]
   {\dimen0=\fontdimen6\the\font 
    \lineskip=#1\dimen0
    \advance\lineskip.5\fontdimen5\the\font
    \advance\lineskip-\dimen0
    \lineskiplimit=.9\lineskip
    \baselineskip=\lineskip
    \advance\baselineskip\dimen0
    \normallineskip\lineskip
    \normallineskiplimit\lineskiplimit
    \normalbaselineskip\baselineskip
    \ignorespaces
   }

\newenvironment{smalldescription}{
   \setlength{\topsep}{0pt}
   \setlength{\partopsep}{0pt}
   \setlength{\parskip}{0pt}
   \begin{description}
   \setlength{\leftmargin}{.2in}
   \setlength{\parsep}{0pt}
   \setlength{\parskip}{0pt}
   \setlength{\itemsep}{0pt}}{\end{description}}

\AtBeginDocument{%
  \providecommand\BibTeX{{%
    \normalfont B\kern-0.5em{\scshape i\kern-0.25em b}\kern-0.8em\TeX}}}

\setcopyright{acmcopyright}
\acmPrice{15.00}
\acmDOI{10.1145/3460319.3464833}
\acmYear{2021}
\copyrightyear{2021}
\acmSubmissionID{issta21main-p85-p}
\acmISBN{978-1-4503-8459-9/21/07}
\acmConference[ISSTA '21]{Proceedings of the 30th ACM SIGSOFT International Symposium on Software Testing and Analysis}{July 11--17, 2021}{Virtual, Denmark}
\acmBooktitle{Proceedings of the 30th ACM SIGSOFT International Symposium on Software Testing and Analysis (ISSTA '21), July 11--17, 2021, Virtual, Denmark}

\begin{document}

\title{HomDroid: Detecting Android Covert Malware by Social-Network Homophily Analysis}

\author{Yueming Wu}
\authornote{Hubei Engineering Research Center on Big Data Security, School of Cyber Science and Engineering, HUST, Wuhan, 430074, China}
\authornote{National Engineering Research Center for Big Data Technology and System, Services Computing Technology and System Lab, HUST, Wuhan, 430074, China}
\affiliation{%
  \institution{Huazhong University of Science and Technology}
  \country{China}
}
\email{wuyueming@hust.edu.cn}

\author{Deqing Zou}
\authornotemark[1]
\authornotemark[2]
\authornote{Shenzhen HUST Research Institute, Shenzhen, 518057, China}
\authornote{Corresponding author}
\affiliation{%
  \institution{Huazhong University of Science and Technology}
  \country{China}
}
\email{deqingzou@hust.edu.cn}

\author{Wei Yang}
\affiliation{%
  \institution{University of Texas at Dallas}
  \country{United States}
  }
  \email{wei.yang@utdallas.edu}

\author{Xiang Li}
\authornotemark[1]
\authornotemark[2]
\affiliation{%
  \institution{Huazhong University of Science and Technology}
  \country{China}
}
\email{xianglilxlx@hust.edu.cn}

\author{Hai Jin}
\authornotemark[2]
\authornote{Cluster and Grid Computing Lab, School of Computer Science and Technology, HUST, Wuhan, 430074, China}
\affiliation{%
  \institution{Huazhong University of Science and Technology}
  \country{China}
}
\email{hjin@hust.edu.cn}

\renewcommand{\abstractname}{ABSTRACT}
\renewcommand{\refname}{REFERENCES}

\begin{abstract}
 
 Android has become the most popular mobile operating system. Correspondingly, an increasing number of Android malware has been developed and spread to steal users' private information. 
 There exists one type of malware whose benign behaviors are developed to camouflage malicious behaviors. The malicious component occupies a small part of the entire code of the application (app for short), and the malicious part is strongly coupled with the benign part.  
 In this case, the malware may cause false negatives when malware detectors extract features from the entire apps to conduct classification because the malicious features of these apps may be hidden among benign features. 
 Moreover, some previous work aims to divide the entire app into several parts to discover the malicious part. However, the premise of these methods to commence app partition is that the connections between the normal part and the malicious part are weak (\eg repackaged malware).
  
 In this paper, we call this type of malware as \emph{Android covert malware} and generate the first dataset of covert malware.
 To detect covert malware samples, we first conduct static analysis to extract the function call graphs.
 Through the deep analysis on call graphs, we observe that although the correlations between the normal part and the malicious part in these graphs are high, the degree of these correlations has a unique range of distribution. 
 Based on the observation, we design a novel system, \emph{HomDroid}, to detect covert malware by analyzing the homophily of call graphs.
We identify the ideal threshold of correlation to distinguish the normal part and the malicious part based on the evaluation results on a dataset of 4,840 benign apps and 3,385 covert malicious apps.
According to our evaluation results, \emph{HomDroid} is capable of detecting 96.8\% of covert malware while the False Negative Rates of another four state-of-the-art systems (\ie \emph{PerDroid}, \emph{Drebin}, \emph{MaMaDroid}, and \emph{IntDroid}) are 30.7\%, 16.3\%, 15.2\%, and 10.4\%, respectively.

%  To detect covert malware, we conduct static analysis to extract the function call graph of it and perform community detection to divide the graph into certain communities (\ie subgraphs).
%  Communities that do not contain any sensitive API calls will be integrated into a \emph{benign community}, and the rest are \emph{sensitive communities}. 
%  For each sensitive community, we perform homophily analysis to obtain the coupling between it and the benign community. 
%  If the coupling is above a predetermined threshold, the sensitive community should be considered benign. 
%  Otherwise, it is treated as a suspicious part of the app. After obtaining all suspicious parts of the app, we integrate them into a subgraph, called \emph{suspicious subgraph}, and conduct statistical analysis to extract the ratio of the number of sensitive triads to the total number of triads within the suspicious subgraph. Finally, these calculated ratios and the occurrences of sensitive API calls are used to detect Android malware.
%  We implement \emph{HomDroid} and conduct evaluations on a dataset of 4,840 benign apps and 3,385 covert malicious apps. 
%  Our evaluation results indicate that \emph{HomDroid} is capable of detecting 96.8\% of covert malware while the False Negative Rates of another three state-of-the-art systems (\ie \emph{PerDroid}, \emph{Drebin}, and \emph{MaMaDroid}) are 30.7\%, 16.3\%, and 15.2\%, respectively.
 
\end{abstract}




\maketitle

\section{INTRODUCTION}

\par The widespread use of smartphones has led to a rapid increase in the number of mobile malware. 
Especially in recent years, mobile phones based on the Android operating system have occupied a dominant position in the smartphone market \cite{report1, report2}, and the Android platform has become the target of choice for attackers due to its huge market share and open-source features. 
To hide the malicious behaviors, attackers have created certain stealthy malware samples~\cite{covertsample1, coversample2, coversample3}. 
For example, \emph{DEFENSOR ID}~\cite{coversample3} removed all potential malicious functionalities but one (\ie abusing Accessibility Service) to hide its maliciousness.
However, as long as this malware is triggered, it can secretly wipe out the victim’s bank account or cryptocurrency wallet and take over their email or social media accounts, which may cause huge economic losses \cite{report3}. 
Therefore, accurate analysis of malware behavior characteristics is urgently needed to remove malware from users' daily life.

\par In general, existing mobile malware detection approaches can be classified into two categories, namely syntax-based approaches \cite{peng2012using, aafer2013droidapiminer, wang2014exploring, arp2014drebin, sarma2012android, zhou2012hey} and semantics-based systems \cite{zhang2014semantics, feng2014apposcopy, mariconti2016mamadroid, yang2015appcontext, allen2018improving, yang2018enmobile, avdiienko2015mining, machiry2018using}. 
As for syntax-based methods, they ignore the program semantics of apps to achieve high efficient Android malware detection. 
For instance, some methods~\cite{peng2012using,wang2014exploring} only consider permissions requested by apps and extract these permissions from manifest to construct feature vectors. 
Nevertheless, malware can spread malicious activities without any permissions \cite{grace2012systematic}. 
To complete more effective Android malware detection, many approaches \cite{gascon2013structural, zhang2014semantics, feng2014apposcopy} distill the program semantics into graph representations and detect malware by analyzing generated graphs.
These graph-based techniques can indeed achieve high accuracy on malware detection. 
For example, \emph{MaMaDroid} \cite{mariconti2016mamadroid} first obtains all sequences from an extracted call graph and then abstracts them into their corresponding packages. 
After abstraction, it establishes a Markov chain model to represent the transition probabilities between these packages. The Markov chain model is used to construct feature vectors to detect Android malware.
The empirical results \cite{mariconti2016mamadroid} have demonstrated the effectiveness of \emph{MaMaDroid} on general malware detection. 
However, when malicious code accounts for a small part of the entire code of an app, the feature vectors obtained from the normal part and the whole call graph can be highly similar.
In this case, \emph{MaMaDroid} will misclassify the malware as a benign app.

% In other words, \emph{MaMaDroid} extracts features from the entire call graph of a given app, thus the similarity of two feature vectors obtained from the normal part and the whole call graph may be very high when the proportion of malicious part is very low (\eg 1\% of the entire call graph).

% The one is the heavy-weight runtime overheads of graph analysis since a graph often contains thousands of nodes. The other is the false negatives induced when detecting certain malware whose malicious code accounts for only a small part of the entire code of an app. 

\par To address the issue, some studies \cite{repack1, repack2} intent to partition the entire app to discover the suspicious part to achieve more accurate malware detection. 
However, these approaches are designed to detect only repackaged Android malware.
A repackaged malware is constructed by disassembling benign apps, adding malicious codes, and then reassembling them as new apps.
In other words, as discussed in their papers \cite{repack1, repack2}, the connections between malicious code and normal code in a repackage malware are expected to be weak because the malicious part is constructed by injection. 
Therefore, if the malicious code is strongly coupled with normal code, methods~\cite{repack1, repack2} may not achieve high accuracy because it is not easy to divide them into different parts.
In short, when the malicious code accounts for a small part (\eg less than 2\%) of a malware sample and the connections between normal code and malicious code are strong, prior approaches \cite{mariconti2016mamadroid, repack1, repack2} may cause high false negatives. 

\par In this paper, we call this type of malware as \emph{Android covert malware} and generate the first covert malware dataset (\ie 3,358 covert malware) by analyzing 100,000 malicious apps' call graphs.
In particular, a malicious sample is considered as a covert malware must meet the following two conditions: 1) Nodes in the malicious part occupy a small part (\eg 2\%) of all nodes in the entire call graph; and 2) The normal part and the malicious part are highly correlated.
Since this type of malware is highly concealed, it may be possible to detect them by performing advanced static analysis or dynamic analysis.
However, the time-cost of these precise analysis methods \cite{yang2015appcontext, yang2018enmobile} are expensive (\eg 291 seconds for \emph{AppContext} \cite{yang2015appcontext} to analyze an app), making it difficult to filter and discover covert malware samples from large-scale real apps.
To tackle the above issues, we aim to propose a novel approach to detect covert malware efficiently and effectively. 
Specifically, we mainly address two challenges:
\begin{itemize}
\item{\emph{Challenge 1: How to identify the malicious code in a covert malware sample?}}
\item{\emph{Challenge 2: How to extract light-weight semantic features to achieve accurate behavior characteristics?}}
\end{itemize}

\par The first key insight of \emph{HomDroid} is the observation from a deep analysis of correlations between the normal part and the malicious part in covert call graphs, that is, although the correlation between the two parts is high, it has a range of distribution.
Specifically, to address the first challenge, we first perform simple static analysis (\ie flow- and context-insensitive analysis) to distill the program semantics of an app into a function call graph. Then the generated graph is divided into certain communities (\ie subgraph) by community detection. 
As malware always performs malicious behaviors by invoking sensitive API calls, therefore, we mainly focus on sensitive API calls in this paper.
In other words, communities that do not contain any sensitive API calls will be integrated into a \emph{benign community}, and the rest are \emph{sensitive communities}.
For each sensitive community, we perform homophily analysis to obtain the coupling between it and the benign community. 
If the coupling is above a predetermined threshold\footnote{In this paper, we select a total of five thresholds: 1, 2, 3, 4, and 5.}, the sensitive community will be considered benign. 
Otherwise, it is treated as a suspicious part of the app. After obtaining all suspicious parts of the app, we integrate them into a subgraph, called \emph{suspicious subgraph}. 

\par The second key insight of \emph{HomDroid} is that triads in social-network-analysis can represent different network structural properties of a network and the extraction of triads from a network is a lightweight task.
Specifically, to address the second challenge, we extract two types of feature sets from the suspicious subgraph. 
We first consider the occurrence of sensitive API calls which are highly correlated with malicious operations.
In addition, to maintain the graph details, we collect the ratio of the number of sensitive triads to the total number of triads within the suspicious subgraph.
By this, we can achieve semantic and efficient Android covert malware detection.

\par We develop an automatic system, \emph{HomDroid}, and evaluate it on a dataset of 8,198 apps including 4,840 benign apps and 3,358 covert malware.
Compared to four state-of-the-art Android malware detection methods (\ie \emph{PerDroid} \cite{wang2014exploring}, \emph{Drebin} \cite{arp2014drebin}, \emph{MaMaDroid} \cite{mariconti2016mamadroid}, and \emph{IntDroid}~\cite{zou2021intdroid}), \emph{HomDroid} is able to detect 96.8\% of covert malware while the False Negative Rates of \emph{PerDroid}, \emph{Drebin}, \emph{MaMaDroid}, and \emph{IntDroid} are 30.7\%, 16.3\%, 15.2\%, and 10.4\%, respectively.
Furthermore, as for runtime overhead of \emph{HomDroid}, it consumes about 13.4 seconds to complete the whole analysis of an app in our dataset.
Such result indicates that \emph{HomDroid} is extremely efficient than methods that perform complex program analysis (\eg 291 seconds for \emph{AppContext} \cite{yang2015appcontext}, 20 minutes for \emph{EnMobile} \cite{yang2018enmobile}, and 275 seconds for \emph{Apposcopy} \cite{feng2014apposcopy}).

\par In summary, this paper makes the following contributions:
\begin{itemize}
\item{We built the first dataset~\cite{HomDroid} of Android covert malware and propose a novel technique to discover the most suspicious part of a covert malware by analyzing the homophily of a call graph.}
\item{We implement a prototype system, \emph{HomDroid}, a novel and automatic system that can accurately detect Android covert malware.}
\item{We conduct evaluations using 4,840 benign samples and 3,358 covert malicious samples. 
Experimental results show that \emph{HomDroid} is capable of detecting Android covert malware with a False Negative Rate of 3.2\% while are 30.7\%, 16.3\%, 15.2\%, and 10.4\% for four comparative systems (\ie  \emph{PerDroid} \cite{wang2014exploring}, \emph{Drebin} \cite{arp2014drebin}, \emph{MaMaDroid} \cite{mariconti2016mamadroid}, and \emph{IntDroid}~\cite{zou2021intdroid})}.
\end{itemize}

\noindent \textbf{Paper organization.} The remainder of the paper is organized as follows. 
Section 2 presents our motivation.
Section 3 introduces our system.
Section 4 reports the experimental results. 
Section 5 discusses the future work and limitations. 
Section 6 describes the related work.
Section 7 concludes the present paper.

\section{MOTIVATION}

\subsection{A Simplified Example}

\par To better illustrate the key insight of our approach, we present a simplified example at first. 
This example (\ie \emph{com.cpssw}) is an app that pushes notifications about the scores of users' favorite teams. 
However, it collects private data such as \emph{International Mobile Equipment Identity} (IMEI), writes them into files, and sends them to a remote server.
The normal behavior of this malware is to bring notifications to users, thus it needs to connect to the Internet to complete this purpose.  
Meanwhile, the malicious activity is to send data to the Internet which can cause privacy leaks. 
In other words, the normal code and malicious code of this malware both need to connect to the Internet to achieve their corresponding goals.

\begin{figure}[htbp]
\centerline{\includegraphics[width=0.48\textwidth]{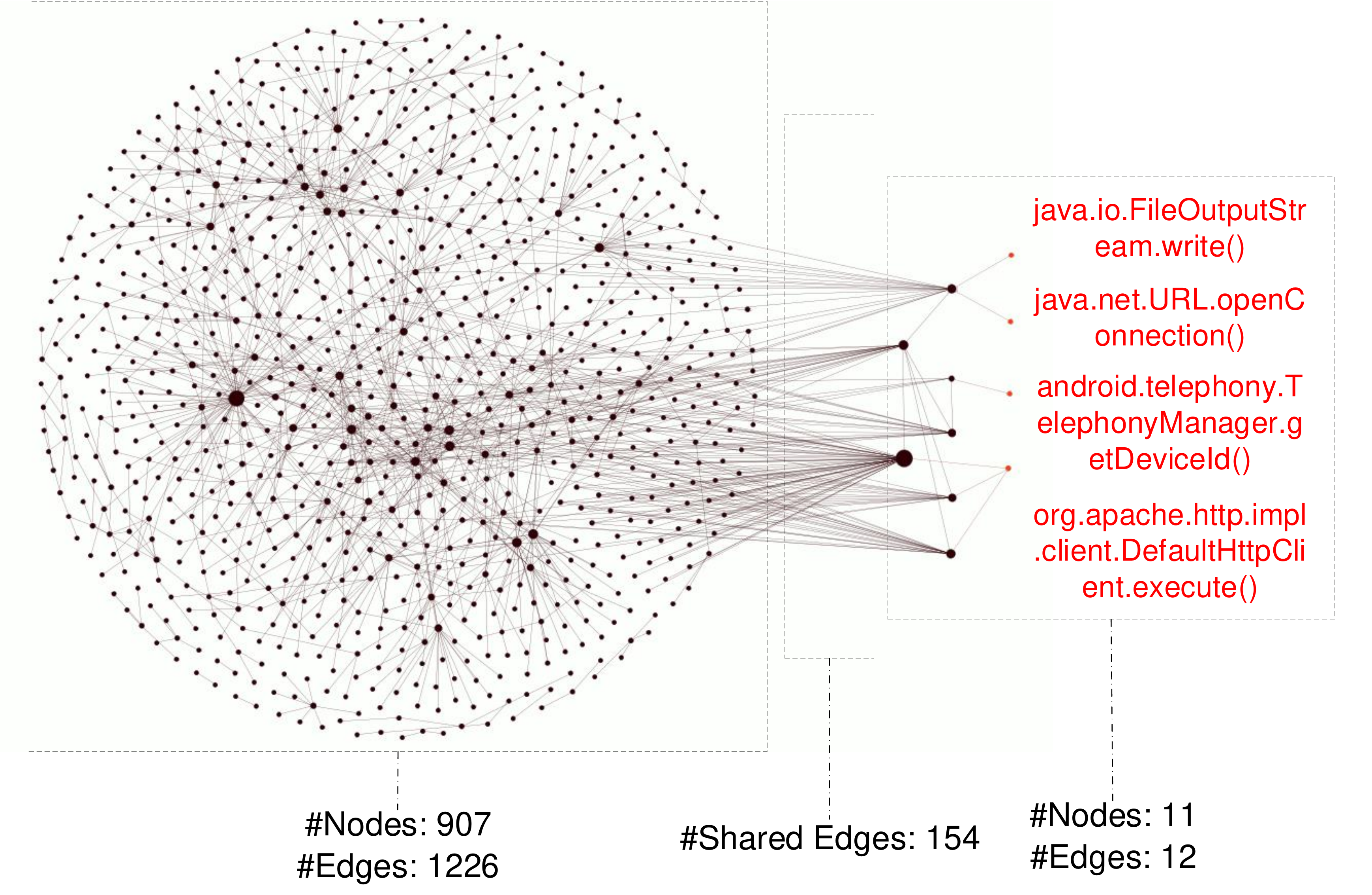}}
\caption{The function call graph of a real malware (com.cpssw)}
\label{fig:moti}
%\vspace{-2em}
\end{figure}

\par Because Android apps use API calls to access operating system functionality and system resources, and malware samples always invoke sensitive API calls to perform malicious activities. 
Therefore, we can leverage sensitive API calls to characterize the malicious behaviors of an app. 
To obtain the sensitive API calls invoked by the malware example, we upload it to an Android apps analysis system \cite{sanddroid} to analyze and generate a detailed behavioral report. 
Through the report, we find that this malware mainly invokes four sensitive API calls (\ie \emph{FileOutputStream.write()}, \emph{URL.openConnection()}, \emph{TelephonyManager.getDeviceId()}, and \emph{DefaultHttpClient.execute()}) to complete its maliciousness. 
In other words, the malicious part of the malware can be characterized by these sensitive API calls and their correlative parent nodes.

% java.io.FileOutputStream.write()
% java.net.URL.openConnection()
% android.telephony.TelephonyManager.getDeviceId()
% org.apache.http.impl.client.DefaultHttpClient.execute()

\par Figure \ref{fig:moti} shows the function call graph of the malware example where the total number of nodes and edges are 918 and 1,392, respectively. 
It consists of two parts, the left part is made up of normal nodes while the right part includes sensitive API calls and their correlative parent nodes. 
As shown in Figure \ref{fig:moti}, the number of nodes of the malicious part is 11 which accounts for about 1.2\% of all nodes of the entire call graph. Moreover, the proportion of edges in the malicious part is also low, which is less than 1\% of all edges of the whole app. 
In this case, we may cause a false negative when we extract features from the entire app to conduct classification since the malicious features may be hidden under normal behaviors.
In practice, the cosine similarity of two feature vectors obtained from the normal part and the whole call graph by \emph{MaMaDroid} \cite{mariconti2016mamadroid} is more than 95\%, and it is misclassified by \emph{MaMaDroid} \cite{mariconti2016mamadroid} in our experiments.

\par Some prior studies \cite{repack1, repack2} intent to partition the entire app to discover the suspicious part to achieve a more accurate malware detection. 
However, these approaches are designed to detect repackaged Android malware and the connections between injected malicious code and legitimate part are expected to be weak. 
In this malware, the number of edges shared by the normal part and malicious part is 154 which is more than 12 times greater than the total number of edges in the malicious part. 
To obtain a more determinate result, we regard the call graph in Figure \ref{fig:moti} as a social network and conduct social-network-analysis to research the homophily of the network.
Homophily is the tendency of individuals to associate and bond with similar others, as in the proverb `birds of a feather flock together' \cite{homophily2001Birds} (Details are in Section 3.3). 
Suppose that a network consists of two subnetworks, high homophily of the network means that the correlation between these two subnetworks is low. 
After analyzing Figure \ref{fig:moti}, we find that the homophily of Figure \ref{fig:moti} is not distinct, in other words, the connections between the normal part and malicious part are strong, making it difficult to be handled by methods in prior studies \cite{repack1, repack2}.
%两端节点不相同的边越多，同质性越不明显，关联性越高。同质性越明显的网络越可以分为多个个体。

\par In conclusion, when the malicious code accounts for a small part of an entire malicious app and the connection between normal code and malicious code is strong, prior approaches \cite{mariconti2016mamadroid, repack1, repack2} may cause high false negatives since the malicious behaviors can be hidden under the normal codes. 
In this paper, we call this type of malware as \emph{Android covert malware} and construct the first dataset of these malware samples.

\subsection{Covert Malware Dataset Construction}

\par To generate the dataset of Android covert malware, we first randomly download about 100,000 malicious apps from AndroZoo \cite{allix2016androzoo} which is a growing collection of Android apps collected from several sources, including the official Google Play app market and several third-party Android app markets (\eg AppChina~\cite{AppChina}). 
After obtaining the samples, we perform static analysis to extract the function call graphs of these apps. 
As API calls are used by the Android apps to access operating system functionality and system resources, they can be used as representations of the behaviors of Android apps. 
Moreover, Android malware usually invokes some sensitive API calls to perform malicious activities. 
Therefore, we assume these sensitive API calls and their correlated parent nodes as the malicious part and the rest as the normal part.
Particularly, we focus on the newest version of API calls set in PScout \cite{au2012pscout}, which is the largest collection of sensitive API calls (\ie 21,986 API calls).

After dividing the graph into the normal part and malicious part, we build the covert malware dataset according to the following steps:
1) We first conduct statistical analysis to obtain the proportion of malicious nodes in all nodes;
2) We then perform homophily analysis to collect the correlation between the normal part and the malicious part;
3) Next, malware samples with low correlation between the normal part and the malicious part are be omitted;
4) As for the remaining malware, we divide them into six categories according to the proportion: [0-1\%), [1\%-2\%), [2\%-3\%), [3\%-4\%), [4\%- 5\%), and greater than 5\%;
5) To find which category may be more covert, we randomly select 100 samples from each category and upload them to VirusTotal~\cite{virustotal};
6) After analyzing the scanning results, we find that malware with a proportion of less than 2\% is less likely to be detected as malware.
Therefore, we pay more attention to these malware samples with a proportion of less than 2\% (\ie 4,321 samples);
7) %In practice, the simple division of the normal part and the malicious part by analyzing sensitive API calls in the call graph may cause some inaccuracies.
To further verify whether these samples contain disguised behaviors or not, we conduct dynamic analysis to obtain accurate behaviors. 
Specifically, we use some state-of-the-art security tools (\eg \emph{AppCritique}~\cite{Appcritique} and \emph{Sandroid}~\cite{sanddroid}) to generate detailed behavior reports, and then manually analyze these reports to check whether sensitive API calls invoked by malicious behaviors are also used by normal behaviors. 
If there are many such cases, then we consider the malware to be covert.
After in-depth manual analysis, we finally obtain 3,358 covert malware samples.

\section{SYSTEM ARCHITECTURE}

\par In this section, we propose a novel system \emph{HomDroid}. We further illustrate how \emph{HomDroid} excavates the most suspicious part of an app and extracts semantic features from the suspicious part to detect Android covert malware.

\begin{figure}[htbp]
\centerline{\includegraphics[width=0.47\textwidth]{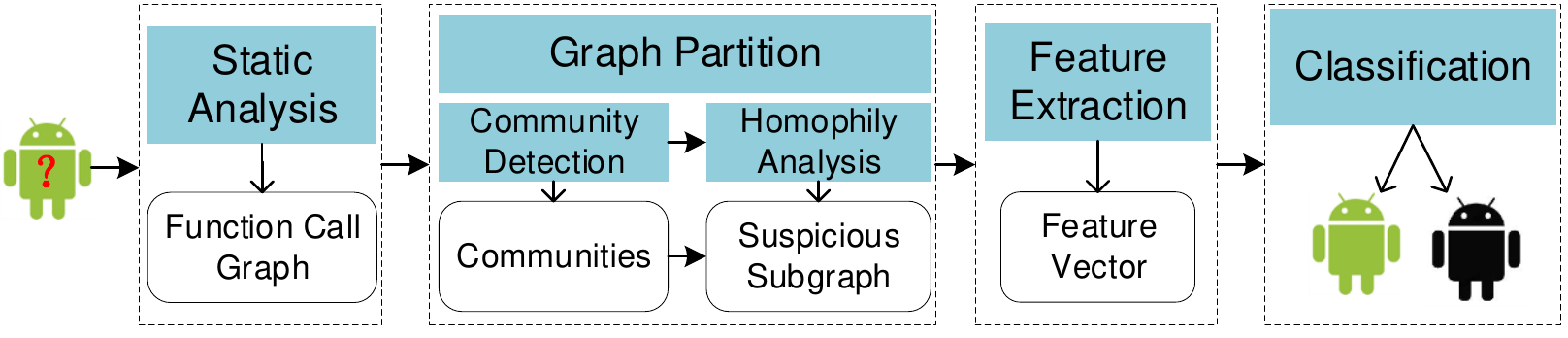}}
\caption{System overview of \emph{HomDroid}}
\label{fig:overview}
%\vspace{-2em}
\end{figure}

\subsection{Overview}
\par As shown in Figure \ref{fig:overview}, \emph{HomDroid} consists of four main phases: \emph{Static Analysis}, \emph{Graph Partition}, \emph{Feature Extraction}, and \emph{Classification}.

\begin{itemize}
\item{\textbf{\emph{Static Analysis:}} Given an app, we first conduct static analysis to obtain the function call graph where each node is a function that can be an API call or a user-defined function.}

\item{\textbf{\emph{Graph Partition:}} After generating the call graph, we then divide it into certain subgraphs by community detection and discover the most suspicious part by homophily analysis. The output of this phase is the most suspicious subgraph.}

\item{\textbf{\emph{Feature Extraction:}} Next, two types of feature sets are collected from the suspicious subgraph, including the appearance of sensitive API calls and the ratio of the number of sensitive triads to the total number of triads within the subgraph.
}

\item{\textbf{\emph{Classification:}} In our final phase, given a feature vector, we can accurately flag it as either benign or malicious by using a trained machine learning classifier.}
\end{itemize}

\subsection{Static Analysis}

To achieve high accuracy malware detection, we consider maintaining the program semantics of an APK file. 
In other words, we conduct light-weight static analysis to distill the semantics of an app into a graph representation. 
More specifically, we implement our static analysis to extract the function call graph of an app based on an Android reverse engineering tool, \emph{Androguard}~\cite{desnos2011androguard}.

% \par To better illustrate the different phases involved in our system, we choose a real-world malware sample\footnote{2af4c588b447963118fd0a8a984438f64898efb0abd01aa6c65dad88d95c7880}. Figure \ref{fig:malwareCG} shows the sample's function call graph, in which, each node is an API call or a user-defined function. The number of nodes and edges are 140 and 251, respectively.

\subsection{Graph Partition}

\par In this phase, we pay attention to discover the most suspicious part of a function call graph. As shown in Figure \ref{fig:overview}, our \emph{Graph Partition} phase is composed of two steps which are \emph{Community Detection} and \emph{Homophily Analysis}.

\subsubsection{Community Detection.}

\par In the study of complex networks, a network is said to have community structure if the nodes of the network can be easily grouped into sets of nodes such that each set of nodes is densely connected internally. 
As for an Android app, it is made up of certain specific modules and each module completes different functionality. 
Nodes in one module should be closely connected because they are designed to implement the same functionality in cooperation. 
Furthermore, a previous study \cite{qu2015exploring} has demonstrated that a software call graph can be treated as a network with community structures. 
Therefore, in this subsection, we perform community detection to divide a function call graph into certain communities (\ie subgraphs).

\par We implement four widely used community detection algorithms (\ie \emph{infomap} \cite{infomap}, \emph{label propagation} \cite{label-propagation}, \emph{multi level} \cite{multi-level}, and \emph{leading eigenvector} \cite{leading-eigenvector}) to check which one is better for us to detect malware.
More specifically, we first download 500 benign apps and 500 malicious apps from AndroZoo \cite{allix2016androzoo} as our test dataset. Then function call graphs of these apps are extracted by our static analysis. 
After obtaining 1,000 call graphs, we conduct community detection on them and record the detection results including the values of modularity \emph{Q} \cite{newman2004finding} of these call graphs and the runtime overheads of different community detection algorithms. 
Modularity \emph{Q} is a metric that can quantify the quality of a detected community structure of a network. The value of \emph{Q} is between 0 to 1, \emph{Q} is 0 means that the network has no community structure. On the contrary, if \emph{Q} is close to 1, the network may have an ideal community structure. 
According to a previous study~\cite{newman2004finding}, the values of \emph{Q} typically fall in the range from about 0.3 to 0.7 and higher values are rare.

\begin{figure}
\centering
\subfigure{
\begin{minipage}[t]{0.22\textwidth}
\centering
\includegraphics[width=\textwidth]{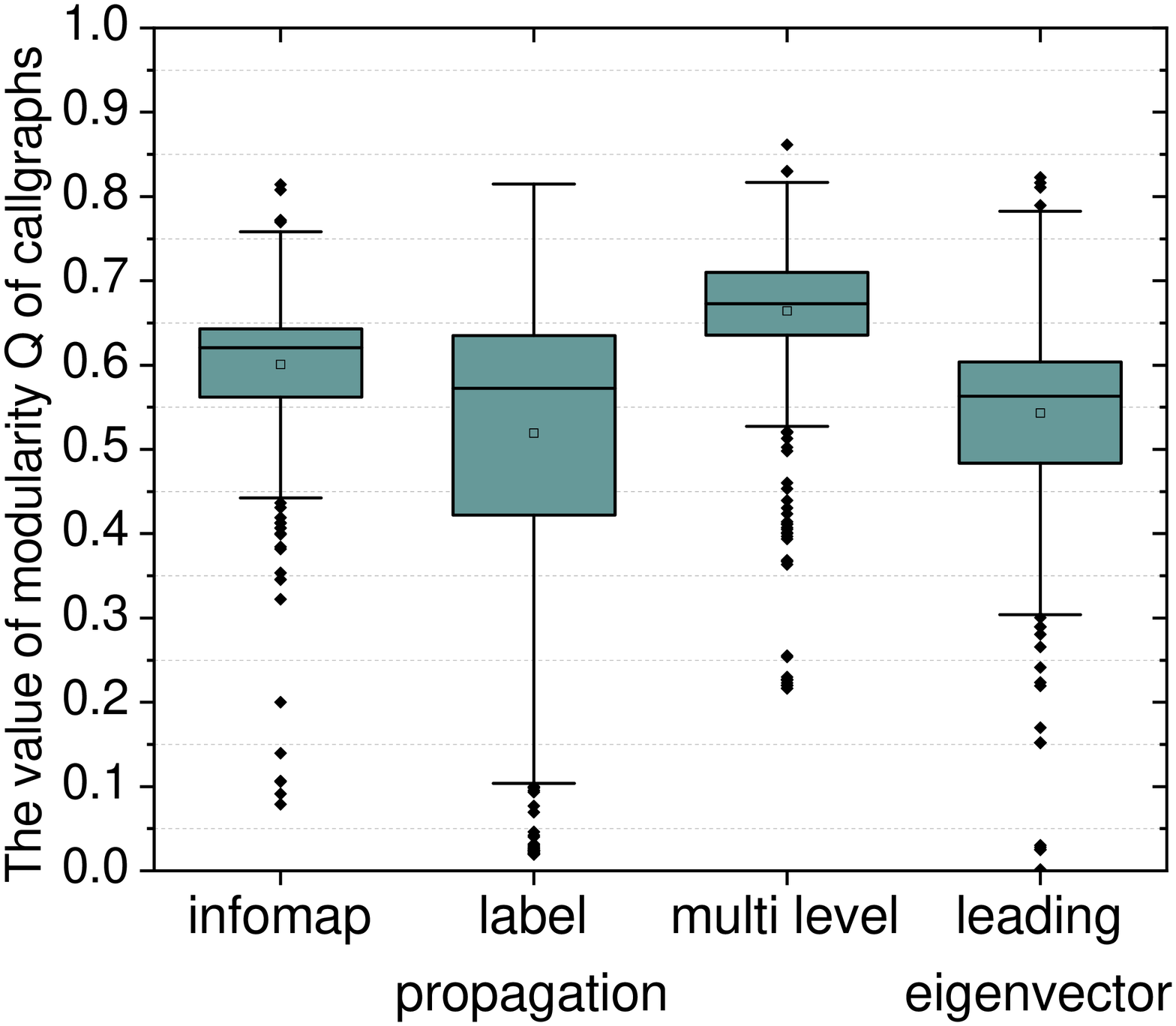}
\end{minipage}
}
\subfigure{
\begin{minipage}[t]{0.22\textwidth}
\centering
\includegraphics[width=\textwidth]{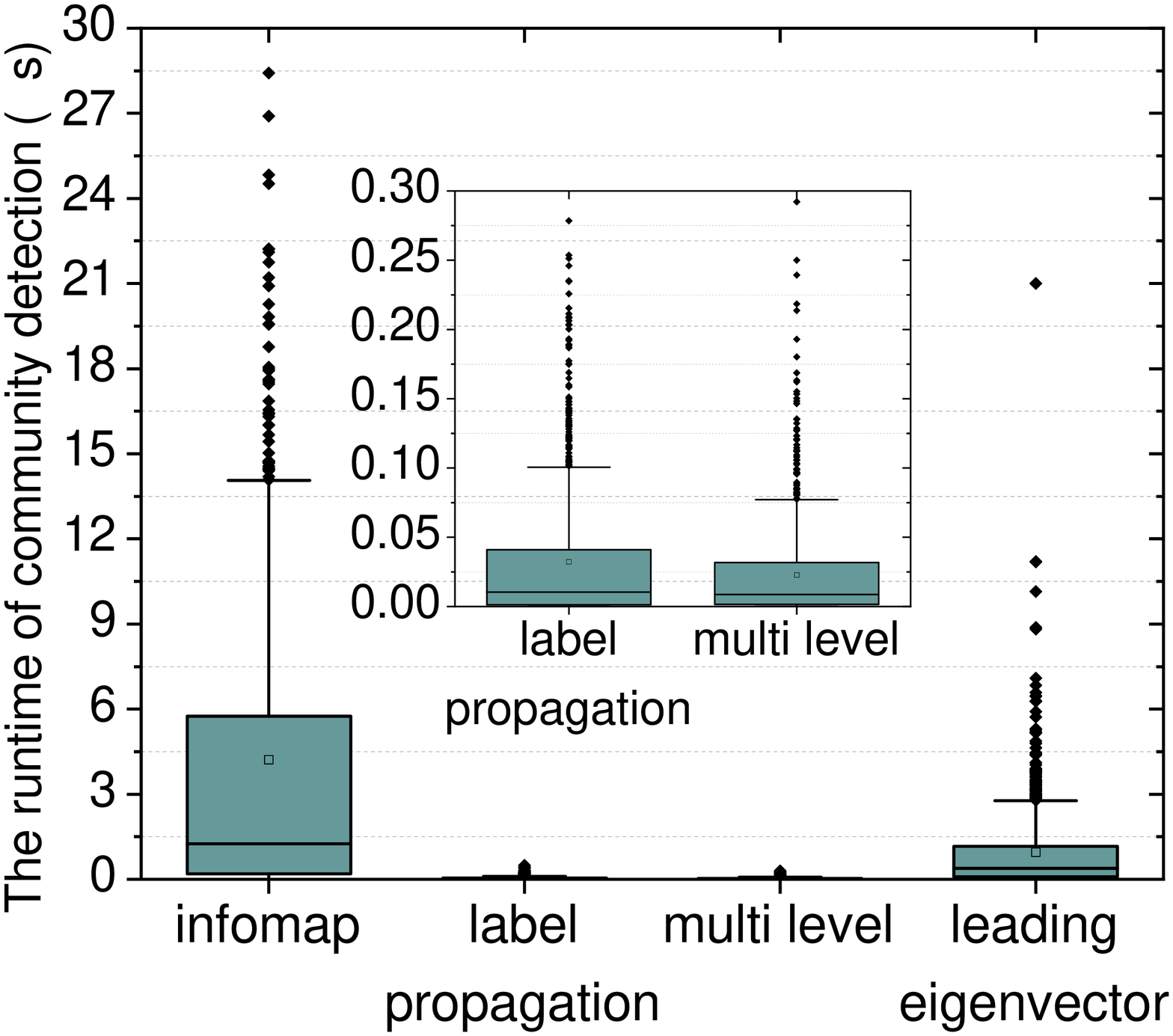}
\end{minipage}
}
\centering
\caption{The values of modularity \emph{Q} of 1,000 call graphs and the runtime overheads of four different community detection algorithms (\ie \emph{infomap} \cite{infomap}, \emph{label propagation} \cite{label-propagation}, \emph{multi level} \cite{multi-level}, and \emph{leading eigenvector} \cite{leading-eigenvector})}
\label{fig:modularity}
\end{figure}

\par Figure \ref{fig:modularity} presents the community detection results on our 1,000 randomly downloaded apps (\ie 500 benign apps and 500 malicious apps). 
On the one hand, the average values of modularity \emph{Q} of communities generated by \emph{informap}, \emph{label propagation}, \emph{multi level}, and \emph{leading eigenvector} are 0.60, 0.52, 0.66, and 0.54, respectively. Such result indicates that communities obtained by \emph{multi level} have better community structure than the other three algorithms. 
On the other hand, as for the runtime of community detection, \emph{multi level} consumes the least runtime overhead which means that it is faster than \emph{informap}, \emph{label propagation}, and \emph{leading eigenvector}.
Therefore, by comprehensively considering the modularity \emph{Q} and the runtime of community detection, we finally decide to choose \emph{multi level} as our community detection algorithm.

\subsubsection{Homophily Analysis.}

\par Homophily is the tendency of individuals to associate and bond with similar others. Homophily has been investigated in many network analysis studies \cite{homophily001, homophily002, homophily003, homophily004}. In this subsection, we use a simple example to better illustrate how to quantify the homophily of a social network.

\begin{figure}[htbp]
\centerline{\includegraphics[width=0.4\textwidth]{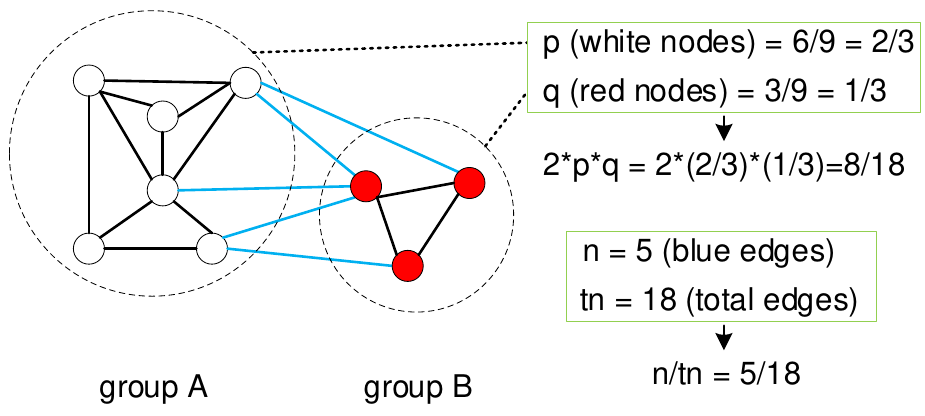}}
\caption{An assumed social network which consists of two groups}
\label{fig:classroom-AB}
%\vspace{-2em}
\end{figure}

Figure \ref{fig:classroom-AB} presents the friendships between students in two groups and connections between them indicate that they are friends. 
For simplicity, we use white nodes and red nodes to represent students in groups A and B, respectively. 
Suppose that the proportions of white nodes and red nodes in Figure \ref{fig:classroom-AB} are \emph{p} and \emph{q}, respectively. 
Because whether the color of a node is red or blue is an independent random process, the probability that a node is white is \emph{p} and the probability of red is \emph{q}. 
Then the probability that two nodes on an edge are different can be computed as \emph{p$*$q} (\ie the color of the left node is white while is red for the right node) + \emph{q$*$p} (\ie the color of the left node is red while is white for the right node) = \emph{2$*$p$*$q}.
Moreover, if the total number of edges in Figure \ref{fig:classroom-AB} is \emph{tn} and the number of edges between groups A and B is \emph{n}. Then the proportion of edges with nodes of different colors can be calculated as \emph{n/tn}.

\par After obtaining the probability that two nodes are different \emph{2$*$p$*$q} and the proportion of edges with two nodes of different colors \emph{n/tn}, we can analyze the homophily of the social network in Figure \ref{fig:classroom-AB} by comparing the value of \emph{2$*$p$*$q} and \emph{n/tn}. In other words, when \emph{n/tn} is less than or equal to \emph{2$*$p$*$q}, the homophily of this social network will be considered high and the correlation between groups A and B is low. On the contrary, if \emph{n/tn} is greater than \emph{2$*$p$*$q}, the social network will have low homophily while the correlation between groups A and B is high.

\par In Figure \ref{fig:classroom-AB}, the proportion of white nodes and red nodes are 2/3 and 1/3, respectively. The number of edges with nodes of different colors is 5 and the total number of edges is 18. In other words, \emph{p}, \emph{q}, \emph{t}, and \emph{tn} in Figure \ref{fig:classroom-AB} are 2/3, 1/3, 5, and 18, respectively. 
Therefore, we can claim that the homophily of the social network in Figure \ref{fig:classroom-AB} is high because 5/18 (\ie \emph{n/tn}) is less than 8/18 (\ie \emph{2$*$p$*$q}). 

\par As stated earlier, the purpose of this phase is to discover the most suspicious part of a given function call graph. 
In reality, Android malware samples usually invoke some sensitive API calls to spread malicious activities. For example, \emph{getDeviceID()} can get your phone’s IMEI and \emph{getLine1Number()} can obtain your phone number. 
To achieve efficient covert malware detection, we only consider a small part of sensitive API calls.
Specifically, we choose sensitive API calls reported in a recent work~\cite{Liangyi2020Experiences} as our objectives which composes of three different API call sets. 
The first API call set is the top 260 API calls with the highest correlation with malware, the second API call set is 112 API calls that relate to restrictive permissions, and the third API call set is 70 API calls that are relevant to sensitive operations. 
In final, 426 sensitive API calls
%\footnote{More detailed information about these sensitive API calls can be found in \cite{Liangyi2020Experiences}.} 
are obtained by computing the union set of these three API call sets. 
However, benign apps may also invoke several sensitive API calls to complete some functionalities (\eg push functionality). 
For instance, some social apps may require to access users' location for presenting location-specific news or videos. 
In the situation they also need to invoke a sensitive API call \emph{LocationManager.getLastLocation()}. 
Therefore, in an effort to achieve more accurate malicious behavior characteristics, we perform homophily analysis to deeply analyze and extract the most suspicious part of a call graph.

Specifically, we define the coupling between two graphs \emph{a} and \emph{b} by computing the quotient of \emph{n/tn} and \emph{2$*$p$*$q}:
\par \centerline{ $c(a, b) = \frac{\frac{s}{e_{a}+e_{b}}}{ 2*\frac{n_{a}}{n_{a}+n_{b}}*\frac{n_{b}}{n_{a}+n_{b}}}$}
 Note that the number of nodes and edges in graph \emph{a} are $n_{a}$ and $e_{a}$ while are $n_{b}$ and $e_{b}$ in graph \emph{b}. Furthermore, the number of edges shared by graph \emph{a} and graph \emph{b} is \emph{s}.

% \begin{algorithm}[htb]
% \small
%   \caption{Discovering the most suspicious subgraph of a function call graph}
%   \begin{algorithmic}[1]
% \REQUIRE
%      $A$: An APK file;
%      $S$: The list of sensitive API calls;
%      $T$: Threshold of relativeness-level.
% \ENSURE
%      $MSS$: The most suspicious subgraph.
     
%  \STATE \textit{CG} $\gets$ \textit{extractCallGraph($A$)} \STATE \textit{Communities} $\gets$ \textit{communityDetection($CG$)}
 
%   \FOR{each $com$ $\in$ $Communities$}
%     \STATE \textit{Nodes} $\gets$ \textit{obtainNodes(com)}
%       \IF{$Nodes$ \& $S$ == $none$}
%          \STATE \textit{BenignCommunities.add(com)}
%       \ELSE
%          \STATE \textit{SensitiveCommunities.add(com)}
%       \ENDIF
%   \ENDFOR
  
%   \STATE \textit{BC} $\gets$ \textit{integrateCommunities(BenignCommunities)}
 
%   \FOR{each $sc$ $\in$ $sensitiveCommunities$}
%     \STATE \textit{coupling} $\gets$ \textit{homophilyAnalysis($BC$, sc)}
%       \IF{coupling $\leq$ $T$}
%          \STATE \textit{SuspiciousCommulities.add(sc)}
%       \ENDIF
%   \ENDFOR
  
%   \STATE \textit{MSS} $\gets$ \textit{integrateCommunities(SuspiciousCommulities)}
   
%   \RETURN{$MSS$}
%   \end{algorithmic}
% \end{algorithm}

%\par Algorithm 1 shows the whole procedure to complete the purpose. 
Given an APK file, we first extract the function call graph by static analysis, then the call graph is divided into certain communities (\ie subgraphs) by community detection.
Communities that do not contain any sensitive API calls will be integrated into a \emph{benign community}, and the rest are \emph{sensitive communities}. 
%This procedure corresponds to Line 3-9 in Algorithm 1.
For each sensitive community, we perform homophily analysis to obtain the coupling between it and the benign community. If the coupling is above a threshold, the sensitive community should be considered benign. 
Otherwise, it is treated as a suspicious part of the app. After obtaining all suspicious parts of the app, we integrate them into a subgraph which is the most suspicious subgraph of the function call graph. 
In this subsection, the purpose of our homophily analysis is to filter the normal parts in sensitive communities (\ie subgraph) to generate a more accurate suspicious subgraph.

\begin{figure}[htbp]
\centerline{\includegraphics[width=0.45\textwidth]{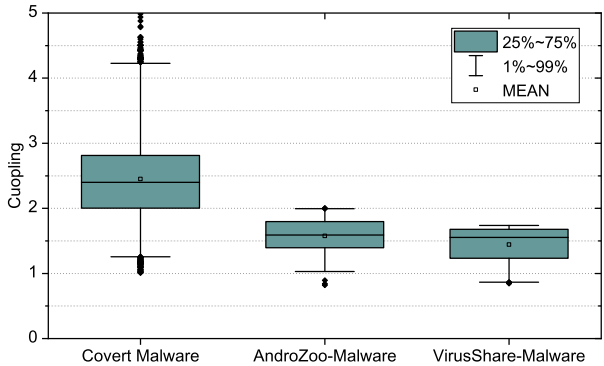}}
\caption{The coupling values between the normal part and the malicious part in Android covert malware and general malware}
\label{fig:level}
%\vspace{-2em}
\end{figure}

In order to select suitable thresholds of coupling, we first conduct a simple study. 
Specifically, we consider 426 sensitive API calls and their correlated nodes as the malicious part and the rest nodes are the normal part.
Then the values of coupling of 3,358 covert malware are extracted.
Moreover, to verify the assumption that the connections between the malicious part and the normal part of covert malware are higher than that of general malware.
We also randomly download 3,000 malware samples from AndroZoo~\cite{allix2016androzoo} and VirusShare~\cite{Virusshare}, respectively.
As aforementioned, we discover 3,358 covert malware from 100,000 malware samples, the percentage is only 3.358\%, which means that most of the malware in our randomly downloaded samples are general malware.
Figure \ref{fig:level} presents the coupling of 3,358 Android covert malware and our randomly downloaded samples from AndroZoo~\cite{allix2016androzoo} and VirusShare~\cite{Virusshare}. 
Through the results in Figure \ref{fig:level}, we observe that the coupling values between the malicious part and the normal part of covert malware are higher than that of most of general malware, and almost all values of coupling in covert malware are between 1 to 5. 
% In other words, although the homophily of these call graphs is low, the correlation between the normal part and malicious part can be quantified (\ie from 1 to 5) to assist the discovery of the most suspicious part of a call graph.
Therefore, we finally choose five thresholds of coupling in this paper, which are 1, 2, 3, 4, and 5.

\begin{figure}[htbp]
\centerline{\includegraphics[width=0.43\textwidth]{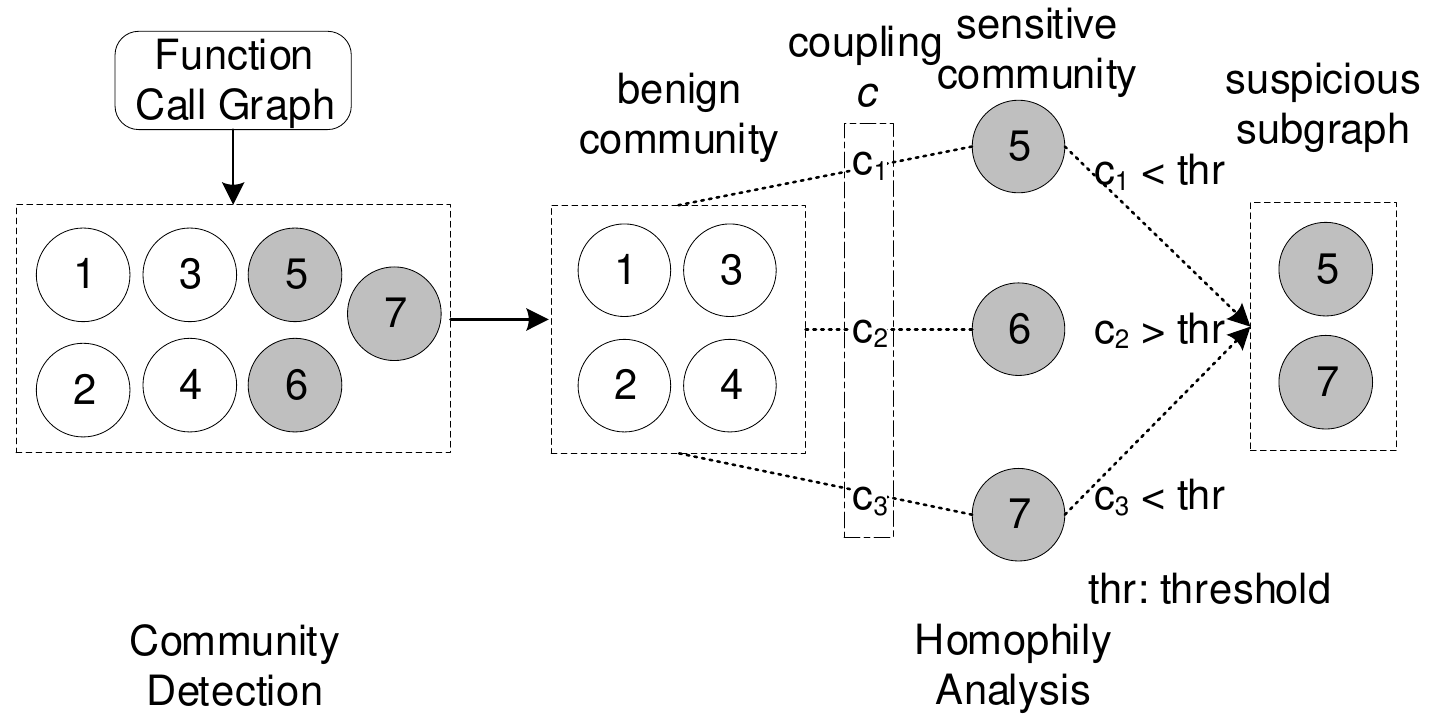}}
\caption{An example of community detection and homophily analysis}
\label{fig:homophily-analysis}
%\vspace{-0.8em}
\end{figure}

\par To better illustrate the different steps involved in community detection and homophily analysis, we present an example in Figure \ref{fig:homophily-analysis}. 
Suppose that a function call graph is partitioned into seven communities and communities 1-4 do not contain any sensitive API calls. 
After generating the benign community by integrating communities 1-4, we perform homophily analysis to compute the coupling between sensitive communities 5-7 and the benign community one by one. 
The analysis results show that the coupling between community 6 and the benign community is less than a given threshold while is greater than the threshold for communities 5 and 7. 
Therefore, we only consider communities 5 and 7 as suspicious parts and integrate them into a subgraph to represent the most suspicious part of the function call graph.

\subsection{Feature Extraction}

\par After obtaining the most suspicious subgraph of a call graph, we then extract features from the subgraph. More specifically, we collect two types of feature sets including the appearance of sensitive API calls and the ratio of the number of sensitive triads to the total number of triads within the subgraph. 

% \begin{figure}[htbp]
% \centerline{\includegraphics[width=0.47\textwidth]{figures/triads-16.jpg}}
% \caption{16 different types of triads in a network.}
% \label{fig:16-triads}
% %\vspace{-2em}
% \end{figure}

\par Our first type of feature set focuses on the occurrence of sensitive API calls. 
As aforementioned, we select 426 most suspicious sensitive API calls \cite{Liangyi2020Experiences} as our concerned objectives.
% As aforementioned, we choose sensitive API calls reported in \cite{Liangyi2020Experiences} as our objectives which composes of three different API call sets. 
% The first API call set is the top 260 API calls with the highest correlation with malware, the second API call set is 112 API calls that relate to restrictive permissions, and the third API call set is 70 API calls that are relevant to sensitive operations. 
% In final, 426 sensitive API calls\footnote{More detailed information about these sensitive API calls can be found in \cite{Liangyi2020Experiences}.} are obtained by computing the union set of these three API call sets. 
Given the most suspicious subgraph, we check whether the nodes in the subgraph contain any sensitive API calls. 
If sensitive API calls appear in the subgraph, the value of the corresponding feature will be one, otherwise, it is zero. 
The dimension of our first type of feature vector is the total number of sensitive API calls, which is 426.

\par Furthermore, to maintain the graph semantics of the subgraph, we also consider the graph structure details to construct our second feature set. 
In social network analysis, there are 16 different types \cite{triads2001A} of triads in a network. 
%which are presented in Figure \ref{fig:16-triads}.
In practice, these triads can represent different network structure properties of a network. 
Therefore, we select a total of six types of triads from the 16 types~\cite{triads2001A} since sensitive API calls are always being invoked by other functions to perform malicious activities in Android malware. 

\begin{figure}[htbp]
\centerline{\includegraphics[width=0.45\textwidth]{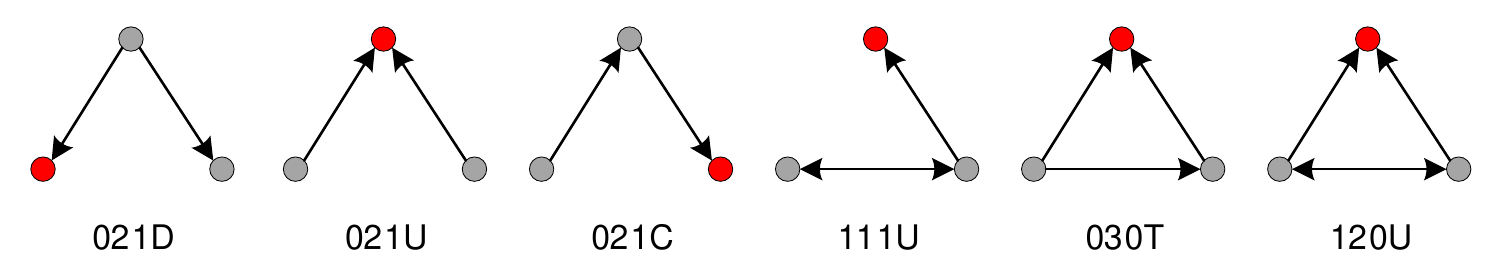}}
\caption{The selected six types of sensitive triads where the red nodes are sensitive API calls}
\label{fig:sensitive-triads}
\end{figure}

\par In Figure \ref{fig:sensitive-triads}, we present our selected six types of triads which can represent different graph structure details. 
For example, a sensitive triad of \emph{021U} shows that two normal functions invoke a same sensitive API call while \emph{021C} indicates that one normal function $f_{1}$ first invokes another normal function $f_{2}$ and then normal function $f_{2}$ invokes a sensitive API call. 
By extracting these sensitive triads from the suspicious subgraph, we are able to characterize malicious behaviors to achieve semantic Android malware detection. 
More specifically, we first extract all these six types of triads from a given suspicious subgraph. 
For each type of triad, we then collect sensitive triads by checking whether a triad contains any sensitive API call or not. 
After obtaining all sensitive triads and all triads, the ratio of the number of sensitive triads to the total number of triads for each sensitive API call will be calculated as the features.  

\begin{figure}[htbp]
\centerline{\includegraphics[width=0.48\textwidth]{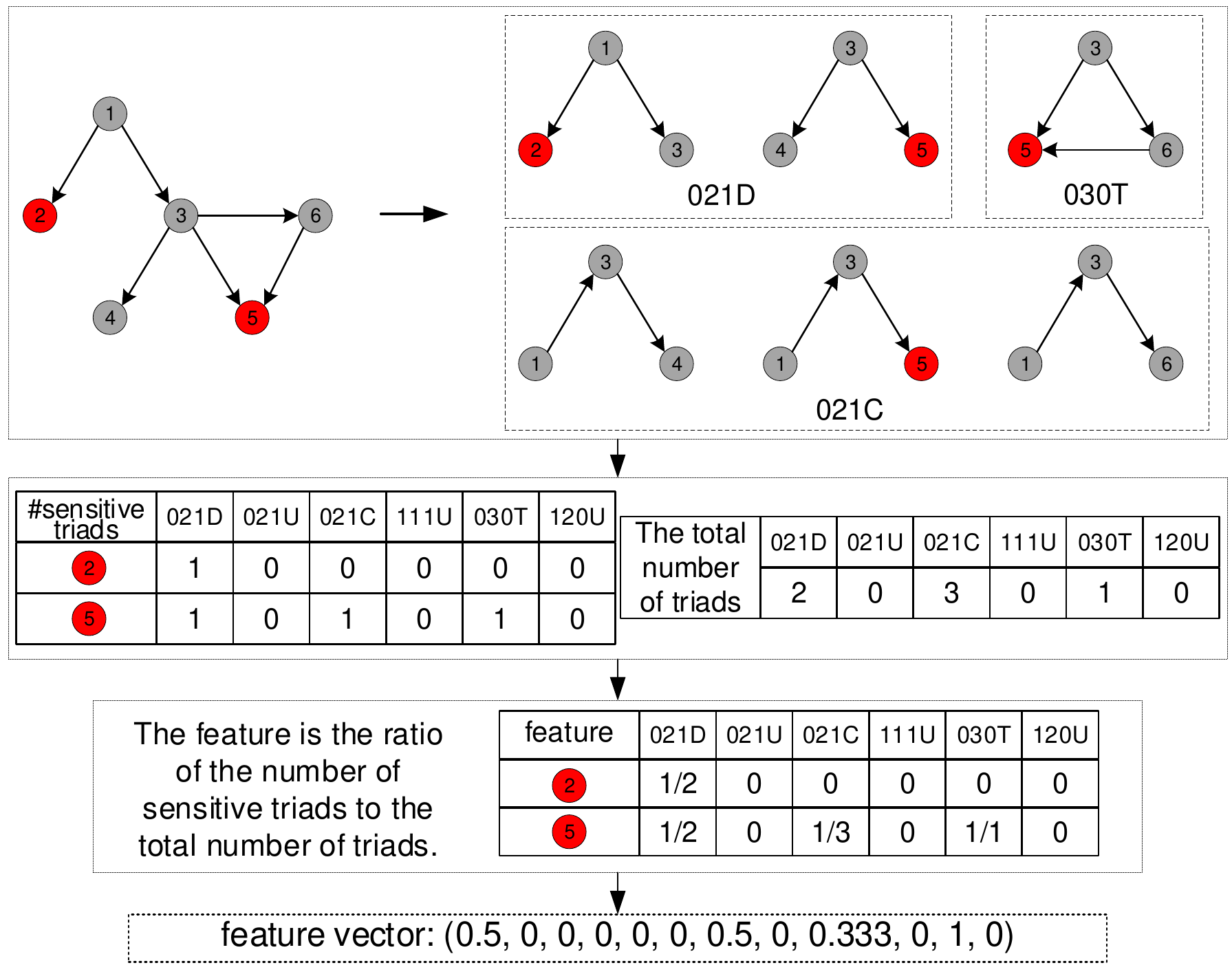}}
\caption{An example to illustrate how we construct the feature vector about the ratio of the number of sensitive triads to the total number of triads}
\label{fig:feature-vector}
\end{figure}

\par To better describe the construction of the features of sensitive triads, we present an example in Figure \ref{fig:feature-vector}. Assuming that a suspicious subgraph contains six nodes, and two of them are sensitive API calls. 
After collecting all the triads from the subgraph, we find that there are four sensitive triads including two sensitive triads of \emph{021D}, one sensitive triad of \emph{021C}, and one sensitive triad of \emph{030T}. 
In addition, the total number of six types (\ie \emph{021D}, \emph{021U}, \emph{021C}, \emph{111U}, \emph{030T}, and \emph{120U}) of triads in the subgraph are two, zero, three, zero, one, and zero, respectively. For each sensitive API call, we compute the ratio of the number of corresponding sensitive triads to the total number of triads for each type of triad. 
For example, the number of triads of \emph{021C} of sensitive API call \textbf{5} is one and the total number of triads of \emph{021C} in the subgraph is three, then the feature of sensitive API call \textbf{5} for \emph{021C} will be calculated as 1/3. Moreover, if the subgraph does not contain sensitive triads of certain types, the corresponding features will be set zero directly. 
Finally, we can obtain a feature vector whose dimension is the number of sensitive API calls multiply by six (\ie six types of sensitive triads). In \emph{HomDroid}, the number of selected sensitive API calls is 426, then the dimension of the feature vector of sensitive triads is 426$*$6=2,556. 

\par After extracting the mentioned two types of features, we concatenate them as our final feature vector whose dimension is 2,982.

\subsection{Classification}

\par Given extracted feature vectors, our final phase focuses on training a classifier first and then uses it to detect Android covert malware. More specifically, in order to check the ability of \emph{HomDroid} on detecting covert malware with different classifiers, 
we implement six classification algorithms (\ie 1-Nearest Neighbor, 3-Nearest Neighbor, Random Forest, Decision Tree, SVM, and Logistic Regression) by using a python library scikit-learn \cite{sklearn}.
Parameters in these algorithms are default parameters in scikit-learn \cite{sklearn}, we leverage them to commence our evaluations and all the experimental results are presented in the following section.

\section{EVALUATIONS}
\par In this section, we aim to answer the following research questions:

\begin{itemize}
  \item \emph{RQ1: How effective is HomDroid on detecting Android covert malware in various setups?}
  \item \emph{RQ2: How does HomDroid perform compared to other state-of-the-art Android malware detection systems?}
  \item \emph{RQ3: What is the runtime overhead of HomDroid on detecting Android malware?}
\end{itemize}

\begin{figure*}
\centering
\subfigure{
\begin{minipage}[t]{0.32\textwidth}
\centering
\includegraphics[width=\textwidth]{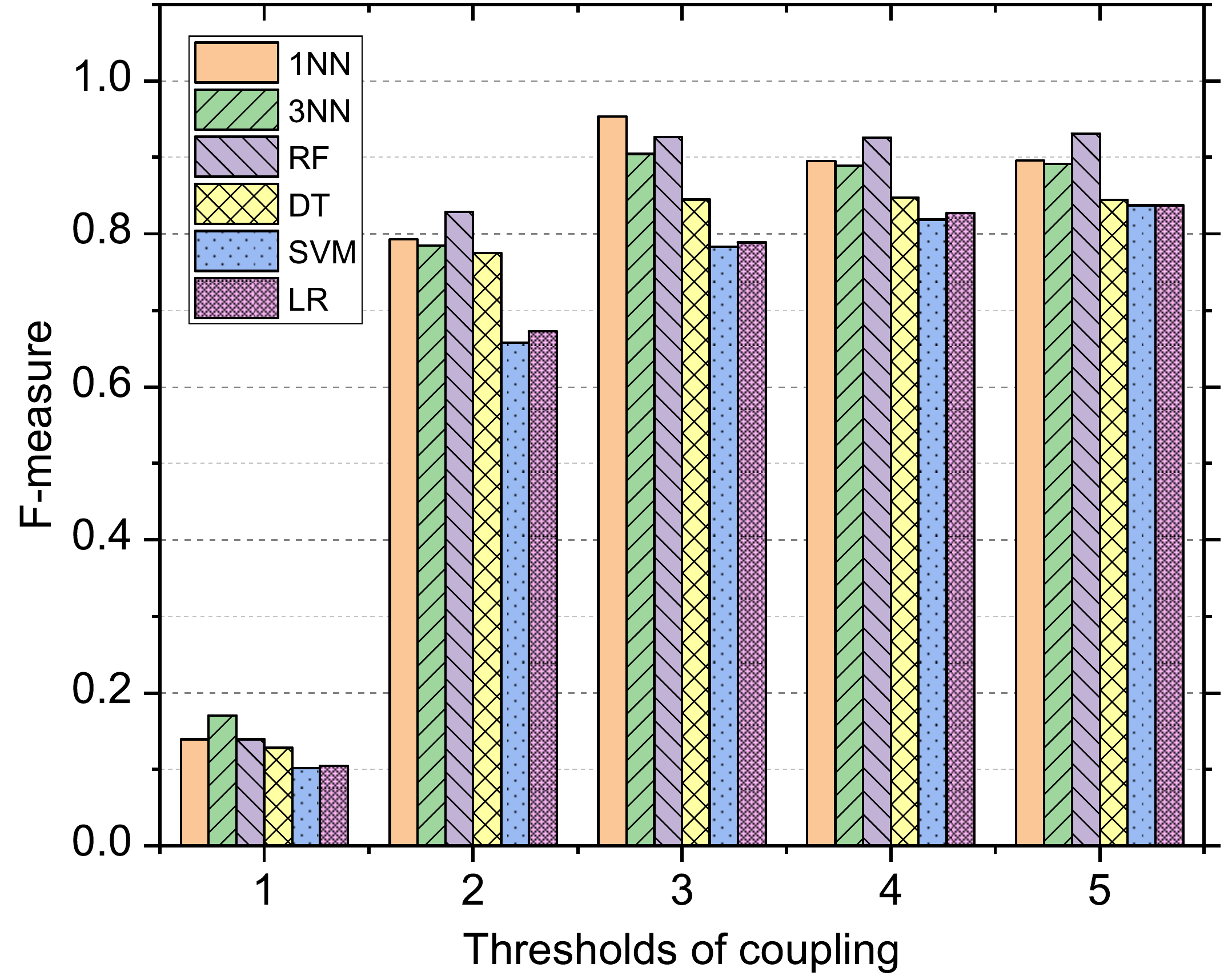}
\end{minipage}
}
% \subfigure{
% \begin{minipage}[t]{0.37\textwidth}
% \centering
% \includegraphics[width=\textwidth]{figures/HomDroid-level5-acc2.pdf}
% \end{minipage}
% }
\subfigure{
\begin{minipage}[t]{0.32\textwidth}
\centering
\includegraphics[width=\textwidth]{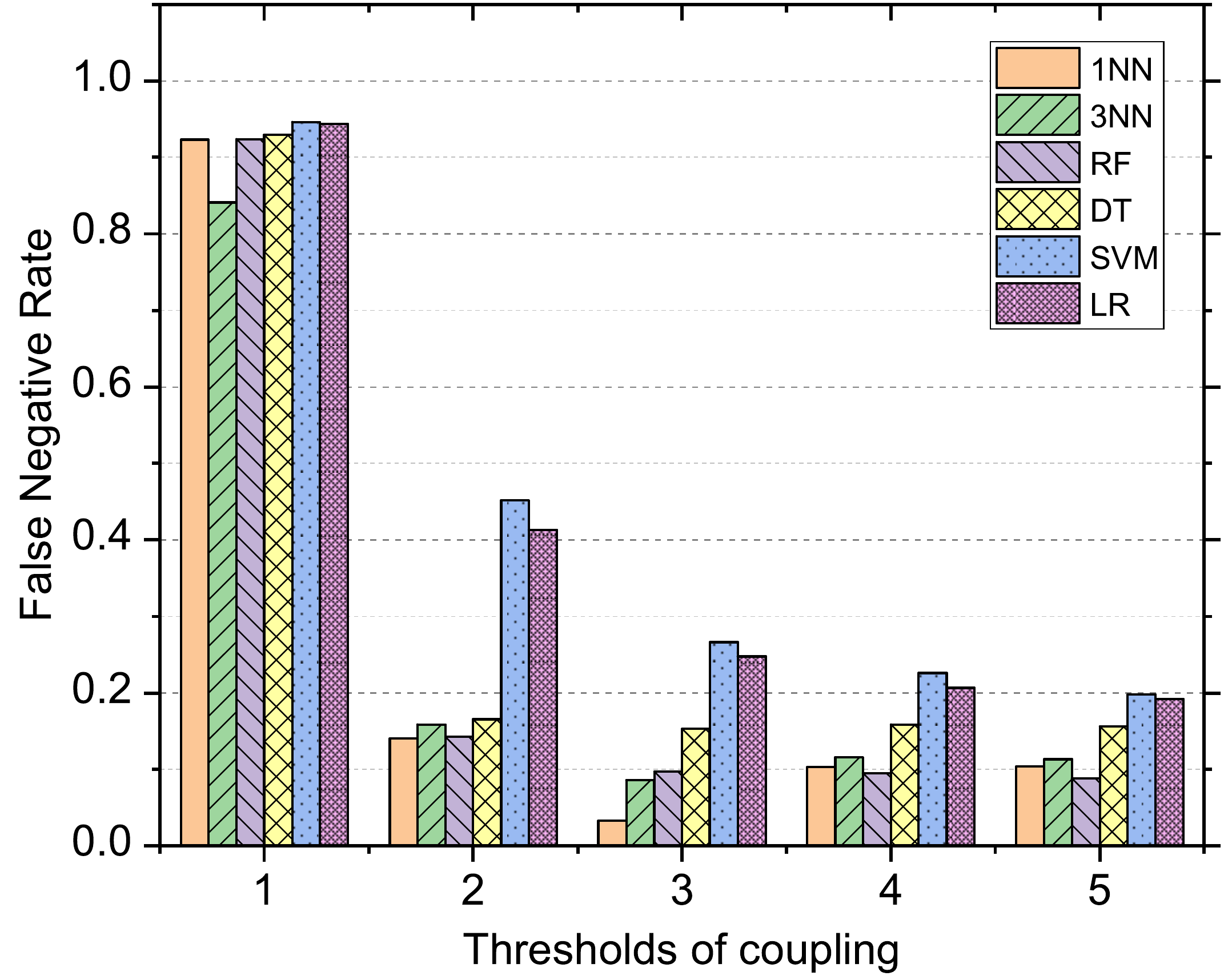}
\end{minipage}
}
\subfigure{
\begin{minipage}[t]{0.32\textwidth}
\centering
\includegraphics[width=\textwidth]{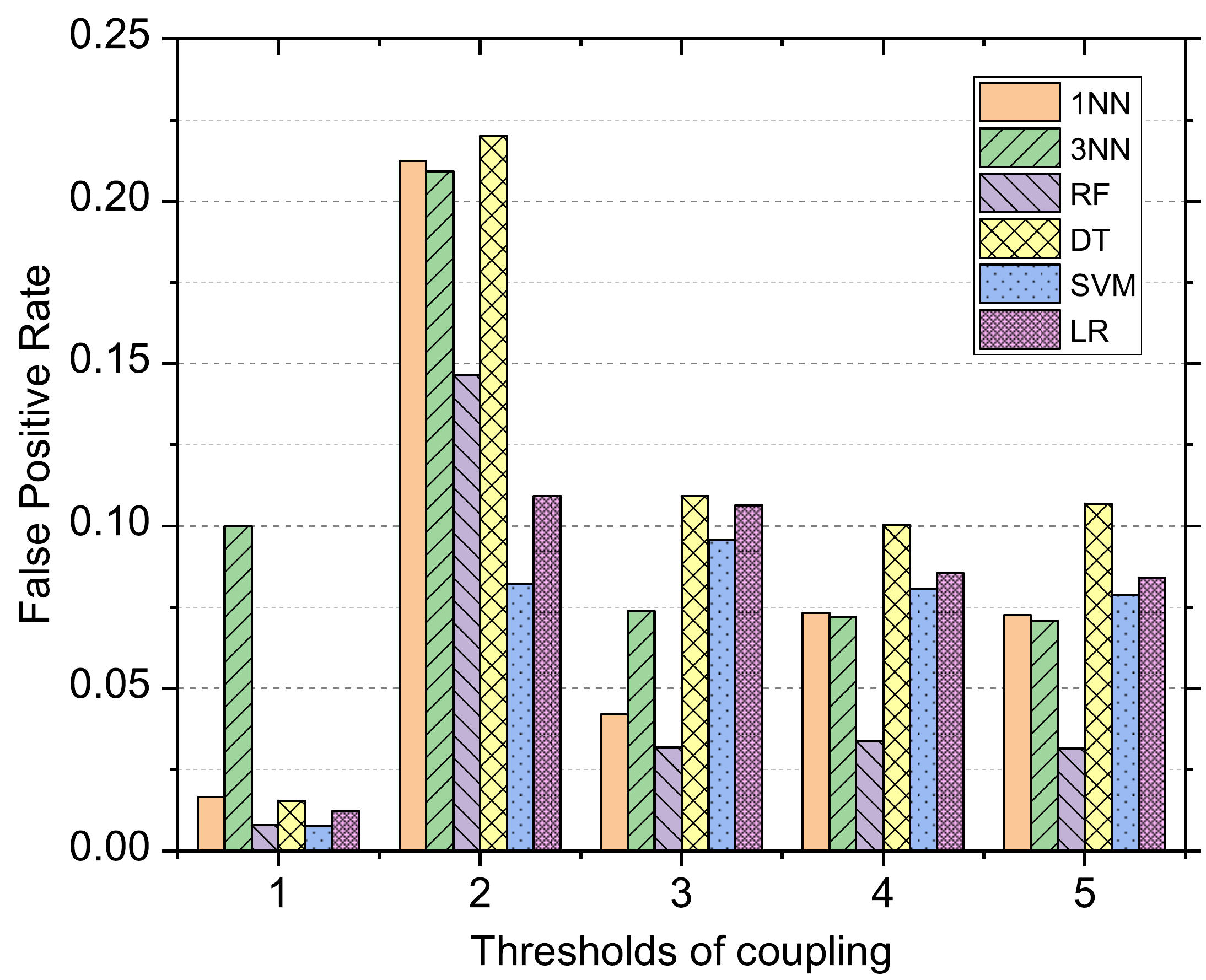}
\end{minipage}
}
\caption{F-measure, FNR, and FPR of \emph{HomDroid} on detecting Android covert malware with different classification models}
\label{fig:f1_tpr_ML}
%\vspace{-1.5em}
\end{figure*}

\subsection{Dataset and Metrics}

\begin{table}[htbp]
  \centering
  \small
    \caption{Summary of the dataset used in our experiments}
   \begin{tabular}{ccc}
   \hline
   \textbf{Category} & \textbf{\#Apps} & \textbf{Average Size (MB)} \\
   \hline
   Benign apps & 4,840 & 2.7 \\
   %\hline
   Covert malware & 3,358 & 3.2 \\
   \hline
   \end{tabular}

   \label{tab:dataset}
  %\vspace{-1.65em}
\end{table}%

%4840+3358=8190
\par As aforementioned, we obtain 3,358 Android covert malware by analyzing and filtering 100,000 malicious apps from AndroZoo \cite{allix2016androzoo}. 
In order to evaluate the effectiveness of \emph{HomDroid} on detecting Android covert malware, we also crawl benign apps from AndroZoo.
Our final dataset consists of 4,840 benign apps and 3,358 covert malware, and the average sizes (in Table \ref{tab:dataset}) of these benign apps and covert malware are 2.7 MB and 3.2 MB, respectively.
We leverage this dataset to commence our evaluations and the experimental results are reported in the following subsections.
Note that all experiments are conducted by performing 10-fold cross-validations, which means that the dataset is partitioned into 10 subsets, each time we pick one subset as our testing set and the rest nine subsets as training set. We repeat this 10 times and report the average as our final results.

\begin{table}
  \centering
  %\footnotesize
  \small
\caption{Descriptions of the used metrics in our experiments}
  \begin{tabular}{m{2.4cm}<{\centering}m{0.6cm}<{\centering}m{4.4cm}<{\centering}}
  %\begin{tabular}{ccc}
  \hline
    \textbf{Metrics}&\textbf{Abbr}&\textbf{Definition}\\
    \hline
    True Positive & \textbf{TP} & \#samples correctly classified as malicious \\
    True Negative & \textbf{TN} & \#samples correctly classified as benign  \\
    False Positive & \textbf{FP} &  \#samples incorrectly classified as malicious \\
    False Negative & \textbf{FN} &  \#samples incorrectly classified as benign \\
    True Positive Rate & \textbf{TPR} & TP/(TP+FN)\\
    False Negative Rate & \textbf{FNR} & FN/(TP+FN)\\
    True Negative Rate & \textbf{TNR}& TN/(TN+FP)\\
    False Positive Rate & \textbf{FPR} & FP/(TN+FP)\\
    Accuracy & \textbf{A} & (TP+TN)/(TP+TN+FP+FN)\\
    Precision & \textbf{P} & TP/(TP+FP)\\
    Recall & \textbf{R} & TP/(TP+FN)\\
    F-measure & \textbf{F1} & 2*P*R/(P+R)\\
    \hline
  \end{tabular}
  \label{tab:metrics}
    %\vspace{-2em}
\end{table}

\par In addition, we adopt several widely used metrics to measure the experimental results of \emph{HomDroid}. 
These metrics are presented in Table \ref{tab:metrics}, in which \emph{True Positive Rate} (TPR) and \emph{False Negative Rate} (FNR) present the effectiveness on detecting malicious samples while \emph{True Negative Rate} (TNR) and \emph{False Positive Rate} (FPR) show the ability on benign samples detection. 
In our experiments, we report both FNR and FPR to see how \emph{HomDroid} performs on classifying both malicious and benign samples.
Moreover, we also report F-measure and Accuracy for presenting the overall detection effectiveness of \emph{HomDroid}.

\subsection{Detection Effectiveness}

\par In this phase, we conduct several experiments to examine the ability of \emph{HomDroid} on Android covert malware detection by using the collected dataset in Table \ref{tab:dataset}. 
Specifically, we evaluate the effectiveness of \emph{HomDroid} from the following two aspects:
\begin{itemize}
  \item  Different classification models: \emph{1-Nearest Neighbor} (1NN), \emph{3-Nearest Neighbor} (3NN), \emph{Random Forest} (RF), \emph{Decision Tree} (DT), \emph{Support Vector Machine} (SVM), and \emph{Logistic Regression} (LR).
  \item Different thresholds of coupling: 1, 2, 3, 4, and 5.
\end{itemize}

\subsubsection{Different Classification Models.}

As aforementioned, we implement six different machine learning algorithms by using a python library scikit-learn \cite{sklearn}. 
These implemented classifiers (\ie 1NN, 3NN, RF, DT, SVM, and LR) are used to evaluate the effectiveness of \emph{HomDroid} on detecting Android covert malware.
Figure \ref{fig:f1_tpr_ML} reports the experimental results including the F-measure, Accuracy, \emph{False Negative Rate} (FNR), and \emph{False Positive Rate} (FPR) achieved by \emph{HomDroid} with different classifiers.
From the results in Figure \ref{fig:f1_tpr_ML}, we can see that classifier 1NN is able to achieve the best overall effectiveness than the other five classifiers. 
For example, when we select the threshold of coupling as 3, the F-measure of \emph{HomDroid} is the highest among all F-measures in Figure \ref{fig:f1_tpr_ML} when we adopt 1NN to detect malware.
The F-measure is 95.3\% while is 90.4\%, 92.6\%, 84.5\%, 78.3\%, and 78.9\% for 3NN, RF, DT, SVM, and LR, respectively.

As for the ability of \emph{HomDroid} on detecting covert malware, the FNR of \emph{HomDroid} is lowest when we choose the threshold of coupling as 3 and 1NN as our final classifier. 
In such case, \emph{HomDroid} only misclassifies about 3.2\% of covert malware as benign apps, which can demonstrate that \emph{HomDroid} has high effectiveness on detecting Android covert malware. 
As regards the effectiveness of \emph{HomDroid} on distinguishing benign samples, RF is able to maintain the lowest false positives than 1NN, 3NN, DT, SVM, and LR no matter which threshold of coupling is selected. 
However, the results of RF are not as good as 1NN in terms of F-measure, Accuracy, and FNR when we choose 3 as the threshold of coupling. 
As a matter of fact, it is crucial to maintain a low FNR for Android malware detection because high FNR signifies more malicious samples being misclassified as benign samples, and these misclassified malicious samples can still spread malicious activities. 
When users install these covert malware samples, their private data may be stolen by attackers, which may cause different levels of economic losses.

\par In conclusion, \emph{HomDroid} can obtain better effectiveness when we adopt 1NN to train a classifier and use it to detect Android covert malware.

% Table generated by Excel2LaTeX from sheet 'Sheet2'
\begin{table}[htbp]
  \centering
  \small
  \caption{F-measure, Accuracy, FNR, and FPR of \emph{HomDroid} on detecting Android covert malware with 1NN}
    \begin{tabular}{c|cccc}
    \hline
    Thresholds & F-measure    & Accuracy & FNR   & FPR \\
    \hline
    1 & 13.9\% & 61.3\% & 92.3\% & 1.7\% \\
    2 & 79.3\% & 81.7\% & 14.1\% & 21.2\% \\
    3 & 95.4\% & 96.2\% & 3.2\% & 4.2\% \\
    4 & 89.5\% & 91.4\% & 10.3\% & 7.3\% \\
    5 & 89.6\% & 91.5\% & 10.4\% & 7.3\% \\
    \hline
    \end{tabular}%
  \label{tab:threshold}%
\end{table}%

\subsubsection{Different Thresholds of Coupling.}

\par In our experiments, we choose five values of thresholds to commence our evaluations according to the result in Figure \ref{fig:level}.
Figure \ref{fig:level} shows that most of the coupling is between 1 to 5, therefore, we select 1, 2, 3, 4, and 5 as our final thresholds to examine the ability of \emph{HomDroid} on covert malware detection.

\par Through the results in Figure \ref{fig:f1_tpr_ML} and Table \ref{tab:threshold}, we can see that \emph{HomDroid} with 1NN is able to maintain the best effectiveness when the threshold of coupling is 3.
In particular, the FNR and FPR of \emph{HomDroid} are only 3.2\% and 4.2\%, which means that about 96.8\% of covert malware and 95.8\% of benign apps can be correctly classified. 
Such result is encouraging and demonstrates that \emph{HomDroid} is able to accurately detect Android covert malware.

\par In short, \emph{HomDroid} can achieve the best effectiveness when we select 3 as our threshold of coupling to generate the most suspicious subgraph and use 1NN to detect covert malware.

\subsection{Comparison with Prior Work}

In this phase, we perform comparative experiments of \emph{HomDroid} with four state-of-the-art Android malware detection approaches: \emph{PerDroid}\footnote{For more convenient discussion, we call the system in \cite{wang2014exploring} as \emph{PerDroid}
since it is a \textbf{per}mission-based method.} \cite{wang2014exploring}, \emph{Drebin} \cite{arp2014drebin}, \emph{MaMaDroid} \cite{mariconti2016mamadroid}, and \emph{IntDroid}~\cite{zou2021intdroid}

\subsubsection{With PerDroid.}

\par \emph{PerDroid} \cite{wang2014exploring} detects Android malware by analyzing risky permissions requested by an app. It scans the manifest file to collect the list of all permissions, and then applies several feature ranking methods to rank them with respect to the risk. 
After obtaining the ranking of all analyzed permissions, permissions with top risks will be considered as risky permissions and are used as features to detect malware. These risky permissions can provide a mechanism of access control to core facilities of the mobile system, thus can be represented as a type of apps' behavior.

\begin{figure}[htbp]
\centerline{\includegraphics[width=0.44\textwidth]{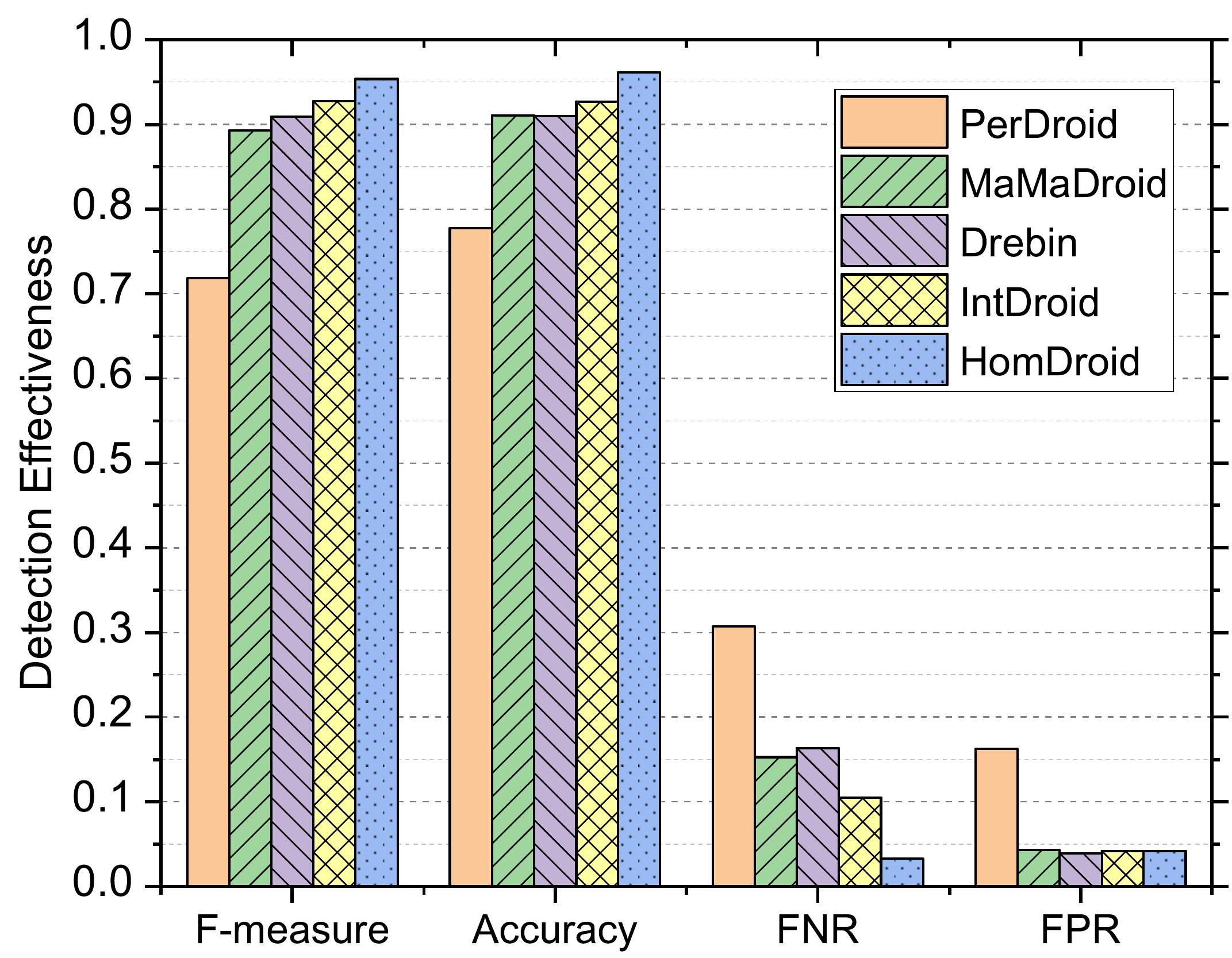}}
\caption{F-measure, Accuracy, FNR, and FPR of \emph{PerDroid}, \emph{MaMaDroid}, \emph{IntDroid}, and \emph{HomDroid} on detecting Android covert malware}
\label{fig:comparative-ml}
\end{figure}

\par The authors \cite{wang2014exploring} have published their risky permissions in their open website \cite{permissionMethod}. Therefore, we directly adopt the list of their published top 88 risky permissions as our features. 
More specifically, we leverage \emph{Androguard} \cite{desnos2011androguard} to implement the feature extraction and scikit-learn \cite{sklearn} to design six classifiers (\ie 1NN, 3NN, RF, DT, SVM, and LR) for completing our malware detection. 
The experimental results on our collected dataset indicate that \emph{PerDroid} is able to maintain the best effectiveness when we choose 3NN as our classifier. 
Therefore, we select the result of 3NN as the effectiveness of \emph{PerDroid} and show the comparative results with \emph{HomDroid} in Figure \ref{fig:comparative-ml}.

% Table generated by Excel2LaTeX from sheet 'Sheet4'
\begin{table}[htbp]
  \centering
  \small
  \caption{FNR of \emph{PerDroid}, \emph{Drebin}, \emph{MaMaDroid}, \emph{IntDroid}, and \emph{HomDroid} on detecting general malware and covert malware (The results of general malware detection are all directly adopted from their published paper \cite{wang2014exploring, arp2014drebin, mariconti2016mamadroid, zou2021intdroid})}
    \begin{tabular}{ccc}
    \hline
    Methods & General Malware & Covert Malware \\
    \hline
    \emph{PerDroid} & 0.054 & 0.307 \\
    \emph{Drebin} & 0.06  & 0.163 \\
    \emph{MaMaDroid} & 0.035 & 0.152 \\
     \emph{IntDroid} & 0.009 & 0.104 \\
    \emph{HomDroid} &   ---    & \textbf{0.032} \\
    \hline
    \end{tabular}%
  \label{tab:comparative-fnr}%
\end{table}%

\par Results in Figure \ref{fig:comparative-ml} indicate that \emph{HomDroid} can detect more covert malware and distinguish more benign apps than \emph{PerDroid}. 
The FNR and FPR of \emph{PerDroid} are 30.7\% and 16.4\% on detecting covert malware while are 3.2\% and 4.2\% for \emph{HomDroid}.
In Table \ref{tab:comparative-fnr}, we also present the FNR of \emph{PerDroid} on general malware detection and the result is directly adopted from their paper \cite{wang2014exploring}.
It is obvious that the effectiveness of \emph{PerDroid} drops a lot when detecting covert malware.
The FNR increases from 5.4\% to 30.7\%, such result indicates that \emph{PerDroid} is not suitable to detect covert malware.
This is mainly due to the fact that \emph{PerDroid} only pays attention to requested permissions and ignores the program details of app code.
However, in order to complete the concealment of malicious behaviors, covert malware samples may use the same permissions as normal code to invoke other sensitive API calls.
Moreover, malware can even perform malicious activities without any permissions \cite{grace2012systematic}. 
On the contrary, \emph{HomDroid} distills the program semantics of app code into a call graph to detect malware, the consideration of program semantics makes \emph{HomDroid} more effective than \emph{PerDroid}.

\subsubsection{With Drebin.}

\par In order to mitigate the inaccuracies in \emph{PerDroid}, \emph{Drebin} \cite{arp2014drebin} has been designed to extract features not only from manifest but also from app code to complete effective Android malware detection. 
More specifically, \emph{Drebin} considers extracting eight different types of feature sets from an app, four of which are extracted from the manifest, and the others are obtained from app code. Feature sets in the manifest consist of hardware features, requested permissions, app components, and filtered intents while includes restricted API calls, used permissions, suspicious API calls, and network addresses in app disassembled code. 
After extracting all the feature sets from an app, \emph{Drebin} embeds them into a joint vector space to train an SVM to detect malware.

%\par We leverage the implementation of \emph{Drebin} in \cite{yang2018enmobile}, and perform comparative experiments on our covert malware dataset. 
Through the experimental results shown in Figure \ref{fig:comparative-ml}, we can see that \emph{Drebin} can achieve better effectiveness than \emph{PerDroid} since \emph{Drebin} considers features from both manifest and app code. 
For example, \emph{Drebin} is able to detect 83.7\% of covert malware while \emph{PerDroid} can only distinguish 69.3\% of these malware samples.
However, although \emph{Drebin} is better than \emph{PerDroid}, it still behinds \emph{HomDroid}. 
It is reasonable because \emph{Drebin} only searches for the presence of certain strings (\ie features) rather than considering the program semantics (\eg invocations between functions). 
In addition, results in Table \ref{tab:comparative-fnr} show that the FNR of \emph{Drebin} increases from 6\% to 16.3\% while is only 3.2\% for \emph{HomDroid} when detecting covert malware. 
Such case indicates that \emph{HomDroid} is superior to \emph{Drebin} on discovering Android covert malware.
This happens because the malicious code of covert malware only accounts for a small part of the entire app, making the features obtained from the benign part and the entire part similar.

\subsubsection{With MaMaDroid.}

\par To complete more comprehensive comparison, we compare \emph{HomDroid} with another graph-based Android malware detection method, namely \emph{MaMaDroid} \cite{mariconti2016mamadroid}. 
Similar to \emph{HomDroid}, \emph{MaMaDroid} first extracts the call graph of an app based on static analysis, then all the sequences are obtained from the call graph and are abstracted into the corresponding packages to model the app's invocation behaviors. 
Specifically, it establishes a Markov chain model to represent the transition probabilities between functions. 
Markov chains stand for multiple pairs of call relationships performed by an app and are used to construct feature vectors to detect malware.

\par The authors \cite{mariconti2016mamadroid} have published their partial source code of \emph{MaMaDroid} \cite{MaMaDroid}. We leverage the open-source code to complete the abstraction, modeling, and feature extraction of \emph{MaMaDroid}. The classification phase is not provided in their code, so we implement the phase by using scikit-learn \cite{sklearn}. 
Specifically, we accomplish the RF classifier to commence our comparative experiments since RF can achieve the best effectiveness as reported in their paper \cite{mariconti2016mamadroid}.

\par Comparative results of \emph{HomDroid} and \emph{MaMaDroid} in terms of F-measure, Accuracy, FNR, and FPR are presented in Figure \ref{fig:comparative-ml}. 
As we can see from Figure \ref{fig:comparative-ml}, \emph{MaMaDroid} outperforms \emph{PerDroid} since \emph{MaMaDroid} maintains the program semantics by distilling them into a call graph. 
However, compared to \emph{HomDroid}, \emph{MaMaDroid} detects less covert malware. 
Specifically, \emph{MaMaDroid} can only detect 84.8\% of covert malware while \emph{HomDroid} is capable of achieving a TPR of 96.8\%. 
In addition, similar to \emph{Drebin}, the effectiveness of \emph{MaMaDroid} also decreases when detecting covert malware, such results are reasonable because of the two following reasons: 
1) The abstraction phase of \emph{MaMaDroid} may incur certain inaccuracies. 
For instance, \emph{android.telephony.TelephonyManager.getDeviceId()} and \emph{android.telephony.SmsManager.sendTextMessage()} are both abstracted in \emph{android.telephony} package while their usages and corresponding security-levels are completely different;
and 2) \emph{MaMaDroid} extracts features from the entire app, which may cause false negatives since the malicious part occupies only a small part of the entire app.

\subsubsection{With IntDroid.}

Our final comparative system is \emph{IntDroid}~\cite{zou2021intdroid}, which combines social-network-analysis-based technique with traditional graph-based method to achieve effective Android malware detection.
Specifically, \emph{IntDroid} first applies centrality analysis to obtain the central nodes in a function call graph, then the average intimacy between sensitive API calls and these central nodes are computed as the semantic features.
To complete our comparative evaluations, we adopt the best parameters in their paper~\cite{zou2021intdroid} to commence our experiments.
In other words, we select nodes with top 3\% all centralities as central nodes and apply 1NN to train a model for malware detection.
The comparative results are shown in Figure \ref{fig:comparative-ml} and Table \ref{tab:comparative-fnr}.

Through the results in Figure \ref{fig:comparative-ml} and Table \ref{tab:comparative-fnr}, we see that \emph{IntDroid} performs better than \emph{PerDroid}.
This happens because \emph{IntDroid} also distills the program semantics into a call graph and then extracts semantic features from the graph.
However, when using \emph{IntDroid} to detect covert malware, the detection effectiveness drops a lot.
For example, \emph{IntDroid} achieves 99.1\% TPR when detecting general malware while the TPR can decrease to 89.6\% when encountering covert malware. 
It is reasonable because the analysis object of \emph{IntDroid} is the entire app, however, the malicious part of covert malware only accounts for a small part of the entire app.
It may cause inaccuracies since the malicious features may be hidden under the benign features.

In conclusion, in distinguishing benign apps, \emph{HomDroid} is similar to other comparison methods because their FPRs are similar.
However, \emph{HomDroid}'s TPR is at least 7\% higher than other comparison tools, which shows that \emph{HomDroid} can find 7\% more covert malware than other tools.

\subsection{Runtime Overhead}

\begin{figure}
\centering
\subfigure{
\begin{minipage}[t]{0.22\textwidth}
\centering
\includegraphics[width=\textwidth]{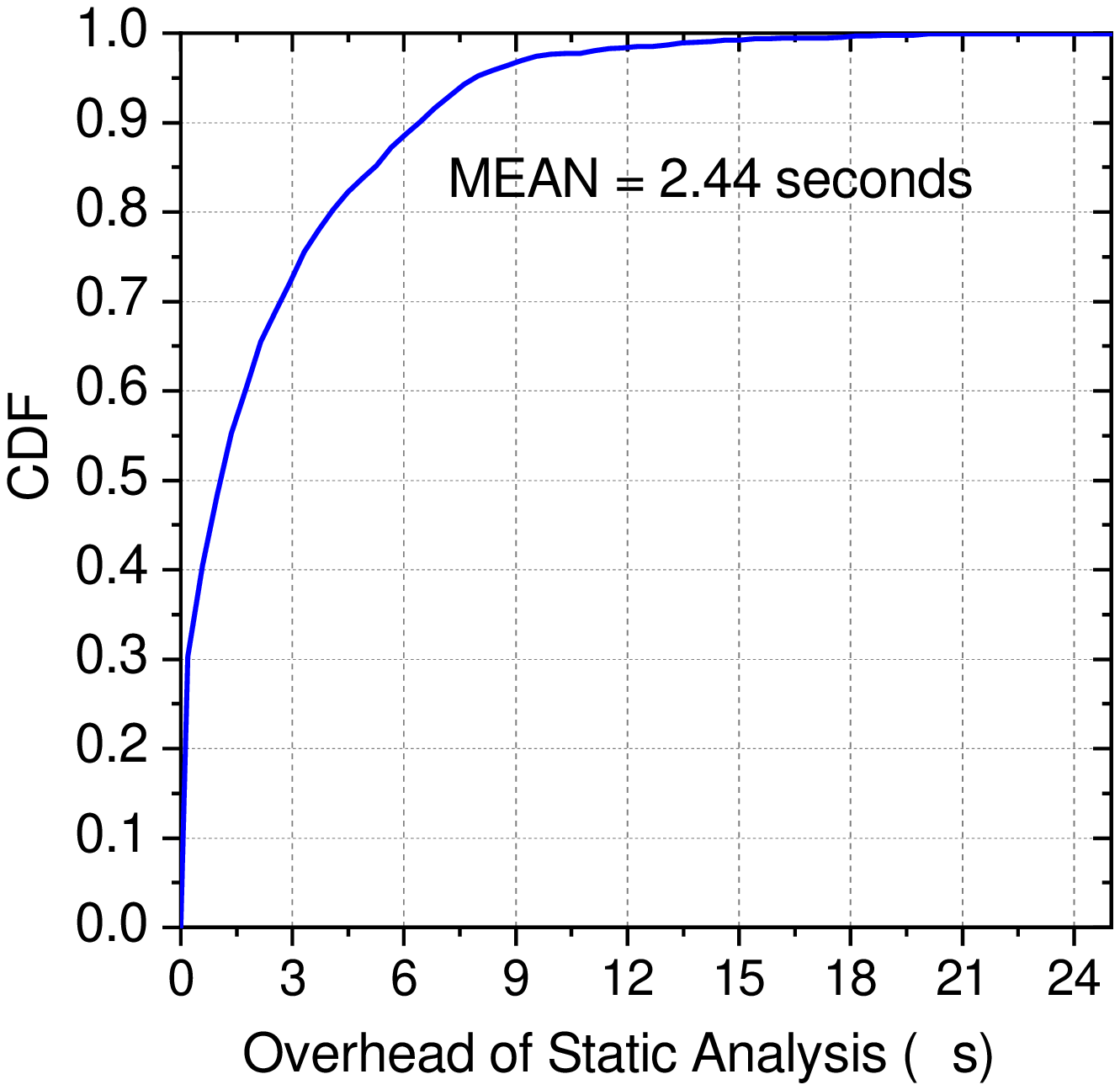}
\end{minipage}
}
\subfigure{
\begin{minipage}[t]{0.22\textwidth}
\centering
\includegraphics[width=\textwidth]{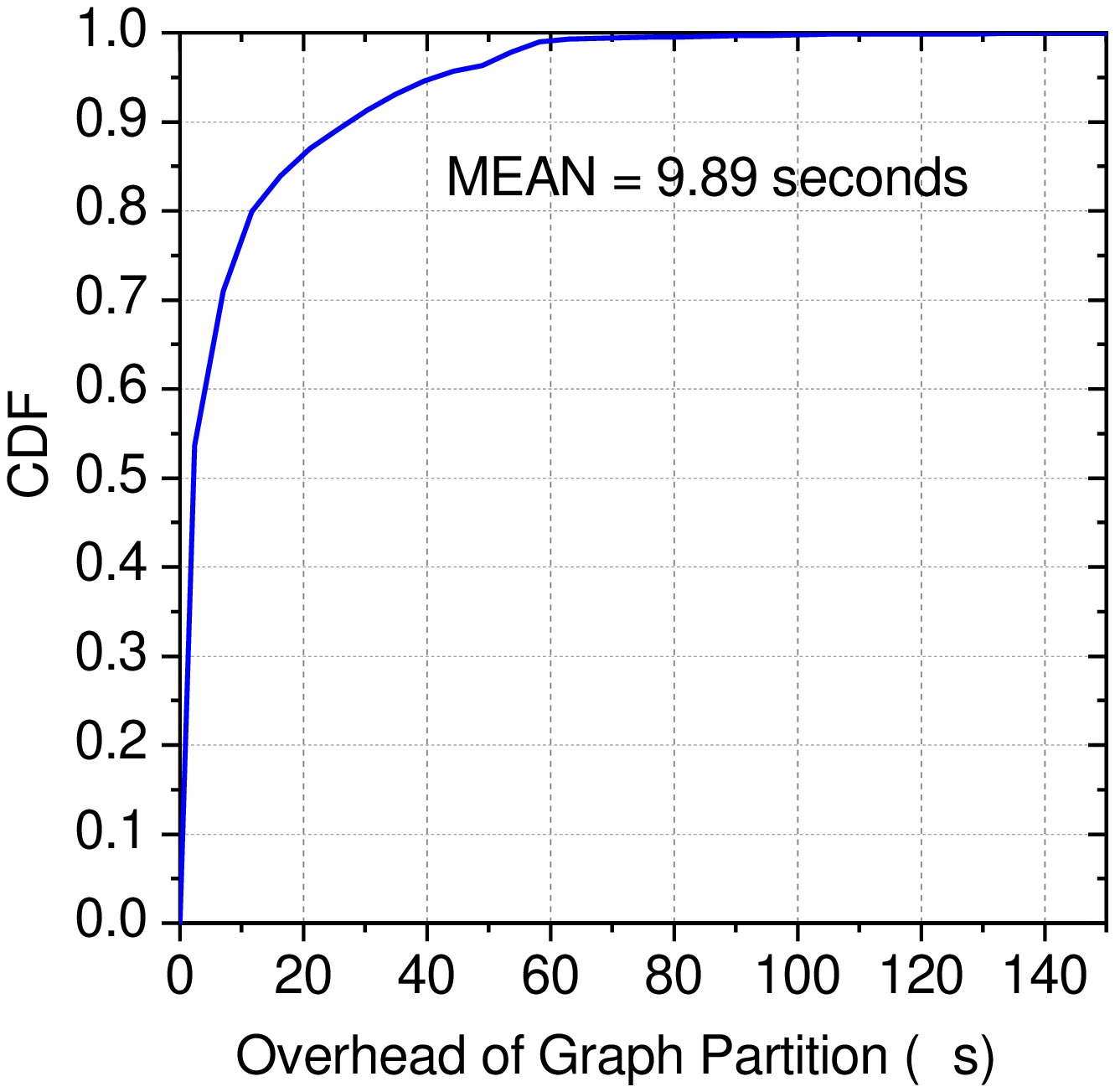}
\end{minipage}
}
\subfigure{
\begin{minipage}[t]{0.22\textwidth}
\centering
\includegraphics[width=\textwidth]{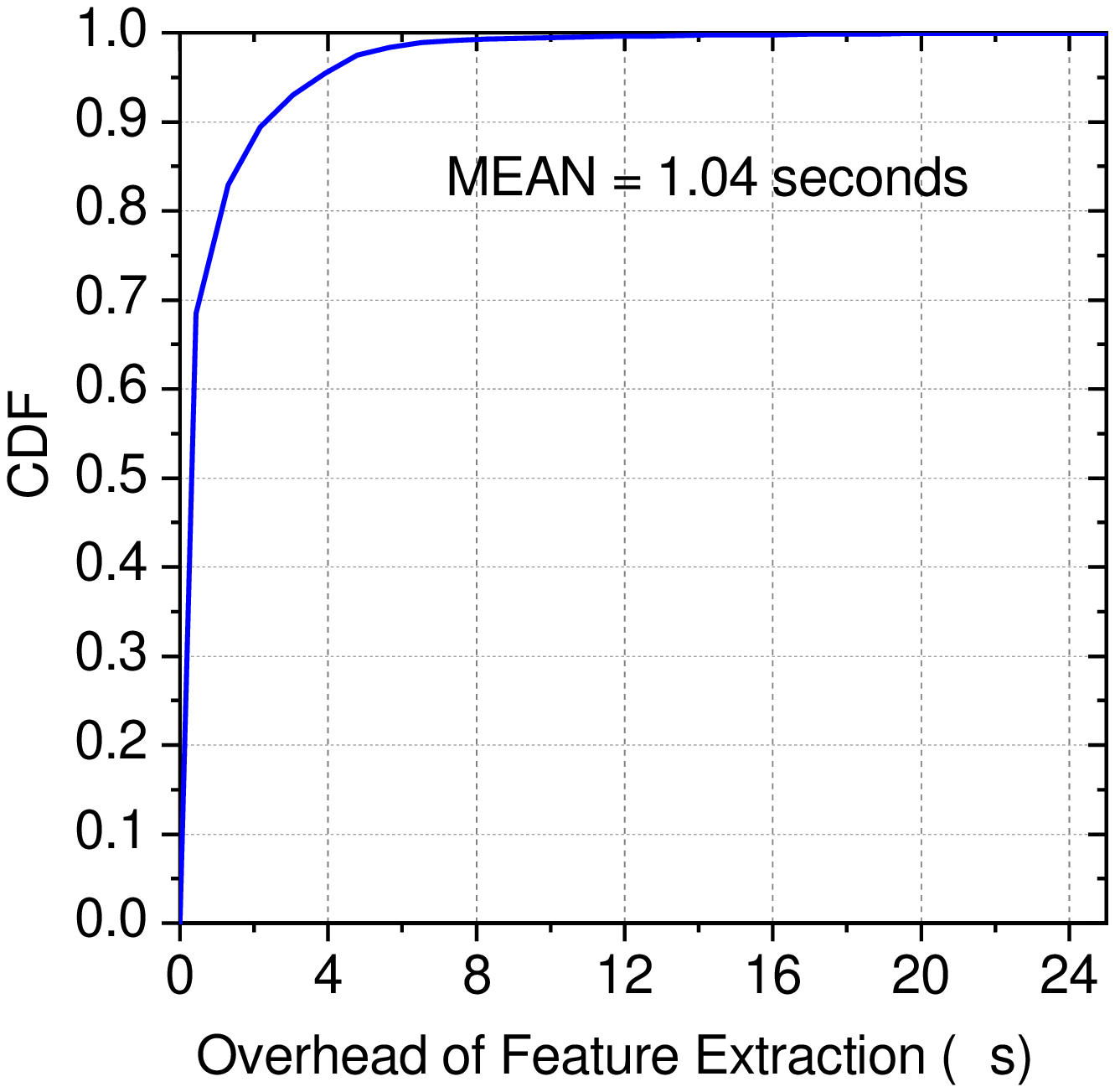}
\end{minipage}
}
\subfigure{
\begin{minipage}[t]{0.225\textwidth}
\centering
\includegraphics[width=\textwidth]{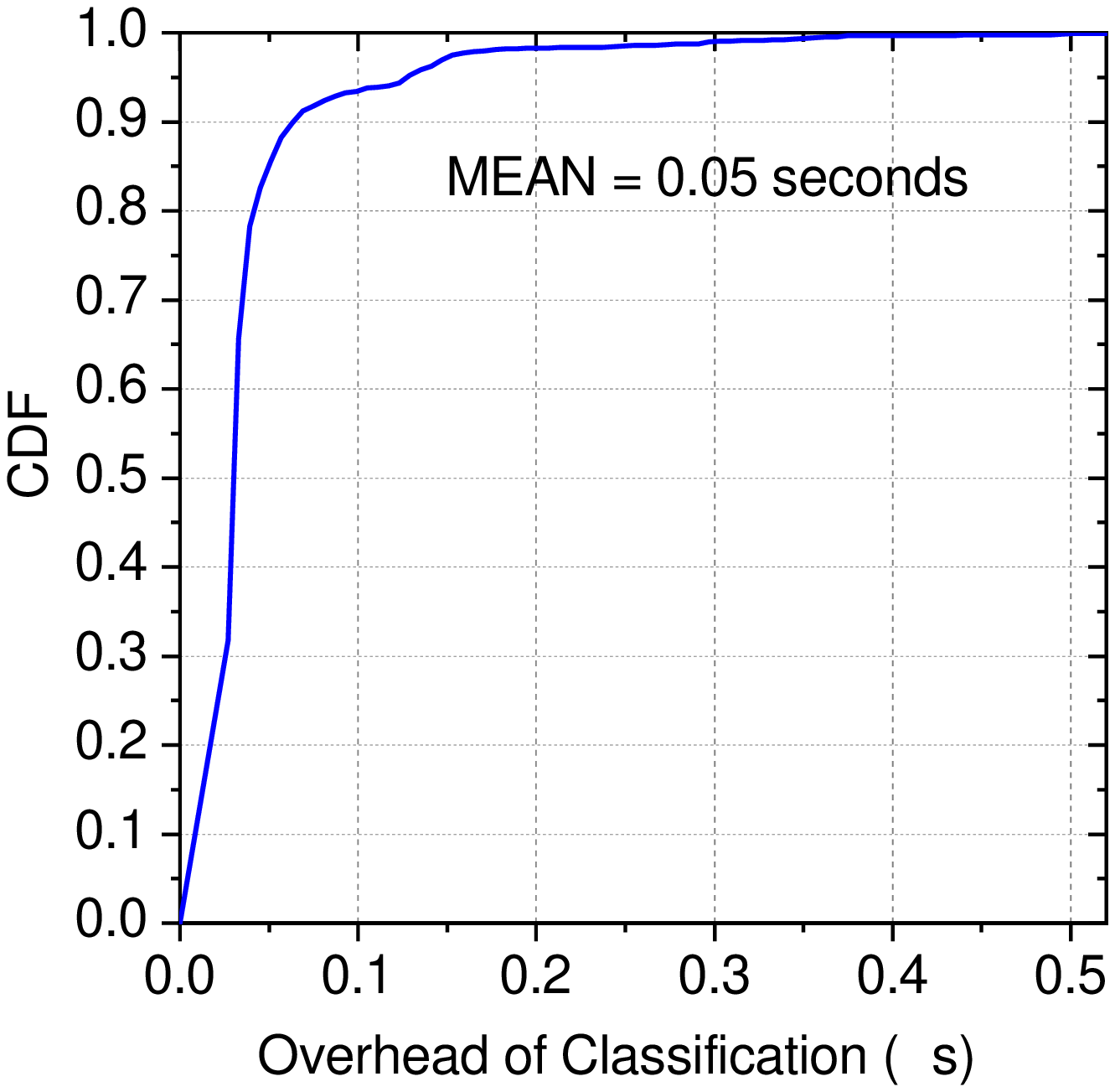}
\end{minipage}
}
\caption{The \emph{Cumulative Distribution Function} (CDF) of runtime overheads of \emph{HomDroid} on different phases (seconds)}
\label{fig:overhead}
%\vspace{-1.5em}
\end{figure}

\par In this phase, we pay attention to evaluate the runtime overhead of \emph{HomDroid}.
As aforementioned, \emph{HomDroid} composes of four steps to analyze an app, which are \emph{Static Analysis}, \emph{Graph Partition}, \emph{Feature Extraction}, and \emph{Classification}. 
We introduce the corresponding overhead in Figure \ref{fig:overhead} according to the different steps involved in \emph{HomDroid}.
Our dataset consists of 4,840 benign apps and 3,385 covert malware, the average nodes and edges of our 8,198 apps are 5,615 and 12,131, respectively.

% Table generated by Excel2LaTeX from sheet 'Sheet2'
% \begin{table}[htbp]
%   \centering
%   \small
%   \caption{The average size of our dataset.}
%     % \begin{tabular}{m{2.3cm}<{\centering}m{1cm}<{\centering}m{2cm}<{\centering}m{2cm}<{\centering}}
%     \begin{tabular}{ccc}
%     \hline
%       Dataset & \#Nodes & \#Edges \\
%       \hline
%     Benign apps  & 3,880  & 8,943 \\
%     Covert malware  & 8,116  & 17,171 \\
%     Total & 5,615  & 12,313 \\
%     \hline
%     \end{tabular}%
%   \label{tab:addlabel}%
% \end{table}%

%benign: 3880 (nodes), 8943 (edges) 
%malware: 8116 + 17171
%average: 5615 + 12313
\emph{(1) Static Analysis}:
\emph{HomDroid} is a semantics-based Android covert malware detection method, it distills the program semantics of an app into a function call graph by static analysis. 
The runtime overhead of function call graph extraction is given in Figure \ref{fig:overhead}, it needs to take about 2.44 seconds to complete the static analysis on average. Moreover, more than 90\% of apps in our dataset can be obtained the call graphs within 8 seconds.

\emph{(2) Graph Partition}:
After obtaining the call graph of an app, we then perform community detection and homophily analysis to dig out the most suspicious subgraph. This step is the most time-consuming phase among all the steps in \emph{HomDroid}. 
As shown in Figure \ref{fig:overhead}, the average runtime overhead to partition a call graph is 9.89 seconds, and more than 85\% of call graphs are able to be discovered the corresponding most suspicious subgraph within 20 seconds. 

\emph{(3) Feature Extraction}:
Given the most suspicious subgraph, we extract two types of feature sets from it including the presence of certain sensitive API calls and the ratio of the number of sensitive triads to the total number of triads within the subgraph. 
This step requires less time than the former two steps (\ie static analysis and graph partition). Specifically, on average, it takes about 1.04 seconds for \emph{HomDroid} to achieve the feature extraction from a suspicious subgraph. 

\emph{(4) Classification}:
The final step of \emph{HomDroid} is to perform classification by a classifier. The classifier is first trained by using feature vectors extracted from the feature extraction step.
Figure \ref{fig:overhead} shows the runtime overhead of \emph{HomDroid} when we select 1NN as our classifier. In practice, classification consumes the least runtime overhead, it only requires about 0.05 seconds to distinguish a feature vector as either benign or malicious.

% Table generated by Excel2LaTeX from sheet 'Sheet3'
\begin{table}[htbp]
  \centering
  \small
  \caption{The comparative runtime overheads of \emph{PerDroid}, \emph{Drebin}, \emph{MaMaDroid}, \emph{IntDroid}, and \emph{HomDroid} on analyzing our datset}
    \begin{tabular}{cc}
    \hline
    Methods & Average Runtime (s) \\
    \hline
    \emph{PerDroid} & 5.67 \\
    \emph{Drebin} & 29.8 \\
    \emph{MaMaDroid} & 60.4 \\
    \emph{IntDroid} & 40.3 \\
    \emph{HomDroid} & 13.4 \\
    \hline
    \end{tabular}%
  \label{tab:comparative-overhead}%
\end{table}%

\par We also compare the runtime overhead of \emph{HomDroid} with \emph{PerDroid}, \emph{Drebin}, \emph{MaMaDroid}, and \emph{IntDroid} as shown in Table \ref{tab:comparative-overhead}.
It is obvious that \emph{PerDroid} is the fastest system to detect malware among these compared four systems since \emph{PerDroid} only extracts the requested permissions from the manifest file of an app.
However, its detection effectiveness is the lowest due to the lack of consideration of semantics from app code. 
As for \emph{Drebin}, it collects eight different types of feature sets from both manifest and app code, and these feature sets consist of several complex features (\eg network address), thus making it time-consuming to complete the whole analysis of an app. 
As regards \emph{MaMaDroid}, due to the complex static analysis and over many features (\ie 115,600 features), the runtime overhead of \emph{MaMaDroid} is more expensive than \emph{Drebin} and \emph{HomDroid}. 
As for \emph{IntDroid}, it combines social network analysis with traditional graph analysis.
Although the runtime overhead of social network analysis is small, the procedure of traditional graph analysis is more time-consuming.
In other words, \emph{HomDroid} achieves a more efficient malware detection than \emph{Drebin}, \emph{MaMaDroid}, and \emph{IntDroid}.

\section{DISCUSSIONS and LIMITATIONS}

\subsection{Discussions}

\emph{\textbf{(1) The use of homophily analysis.}}
In our work, the generation of covert malware should meet that the normal parts and the malicious parts are highly correlated.
In this generation phase, the use of homophily analysis is to check whether the normal parts and the malicious parts are highly correlated or not.
Besides, in \emph{graph partition} phase of \emph{HomDroid}, the purpose of homophily analysis is to filter out the normal parts from sensitive subgraphs to generate an accurate suspicious subgraph. 
Although we choose to leverage homophily analysis in both phases, the correlation analysis in the two phases are independent. The use of same analysis technique will not incur bias in the final results.

\emph{\textbf{(2) Why can HomDroid detect covert malware?}}
Existing Android malware detection methods analyze the entire app to extract features, however, the malicious part of covert malware only occupies a small part of the entire app, thus the malicious features may be hidden under benign features. 
Moreover, some repackaged malware detection techniques first divide the entire app into several parts and then discover the most suspicious part. 
The premise of these methods to commence app partition is that the connections between the normal part and the malicious part are weak. 
However, covert malware samples do not fit the situation.
\emph{HomDroid} performs homophily analysis to discover the most suspicious parts, which has the ability to discover the suspicious parts from the entire graph although the connections between the normal part and the malicious part are strong.

\emph{\textbf{(3) The application of HomDroid.}}
The design of \emph{HomDroid} is to detect Android covert malware, therefore, we can combine \emph{HomDroid} with other state-of-the-art general malware detection techniques (\eg \emph{IntDroid}) to complete more comprehensive malware discovery.
For example, \emph{IntDroid} can be used as the first line of defense to filter most of malware (\ie general malware), then \emph{HomDroid} can be applied as the second line of defense to discover more malware (\ie covert malware).

\emph{\textbf{(4) Using different triads to detect covert malware.}}
Since sensitive API calls are always being invoked by other functions to perform malicious activities, therefore, we only select six types of triads to commence our feature extraction.
In reality, we also apply a feature ranking method (\ie T-test) to research the ability of these six triads on covert malware detection.
The results show that \emph{021U} triad ranks first.
In our future work, we will construct different combinations of triads to find the most suitable combination to achieve better detection results.

\subsection{Limitations}

\par Similar to any empirical approach, \emph{HomDroid} suffers from several limitations, which are listed below.

\emph{\textbf{(1) Call graph extraction.}} 
In this paper, our static analysis phase is implemented by leveraging \emph{Androguard} \cite{desnos2011androguard}.
In reality, function call graph extracted by \emph{Androguard} \cite{desnos2011androguard} is a context- and flow-insensitive call graph. 
We ignore these information for achieving high efficiency to detect covert malware.
To mitigate the inaccuracies caused by our constructed call graph, we plan to use advanced program analysis to generate a suitable call graph to achieve the balance between the efficiency and effectiveness on detecting covert malware.

\emph{\textbf{(2) Obfuscations.}}
Since \emph{HomDroid} is a graph-based method, it can resist some typical obfuscations (\eg rename obfuscation), however, apps can use reflection \cite{rastogi2013catch} to call sensitive API calls, in this case, we may miss the call relationships between these methods.
To be resilient to reflection, we can use an open-source tool, \emph{DroidRA}~\cite{li2016droidra}, to conduct reflection analysis to identify methods that use reflection for each app. 
Then the missing edges can be added into the call graph, where caller nodes are methods that use reflection and callee nodes are reflected methods.
Moreover, since \emph{HomDroid} is to perform static analysis to extract the call graph, it is vulnerable to dynamic loading and encryption (\eg \emph{APK Protect} \cite{apkprotect}). 
As for dynamic loading, it is the technique through which a computer program at runtime loads a library into memory.
Thus all static-analysis-based methods suffer from this limitation.
As for encryption, packers can protect apps by using encryption techniques to hide the actual Dex code. 
To address this limitation, we can use some unpacker systems such as \emph{PackerGrind}~\cite{xue2017adaptive} to recover the actual Dex files, then static analysis can be applied to extract call graph.

\emph{\textbf{(3) Sensitive API calls.}} \emph{HomDroid} mainly focuses on 426 sensitive API calls that are highly correlated with malicious operations \cite{Liangyi2020Experiences}.
These sensitive API calls account for a small part of the whole sensitive API calls. 
We plan to conduct statistical analysis to select more valuable sensitive API calls to commence our experiments.

\emph{\textbf{(4) The partition of our dataset.}}
We perform 10-fold cross-validations to generate our evaluation results, which means that the dataset is partitioned into 10 subsets, each time we pick one subset as our testing set and the rest nine subsets as training set. 
We repeat this 10 times and report the average.
In this partitioning phase, we do not distinguish the family labels of our covert malware.
In practice, the result may overfit when samples from the same family are not used in both training and testing sets.
We plan to label the family first by using \emph{Avclass}~\cite{sebastian2016avclass}, and then make sure that samples from the same family are used in both training and testing sets.

\section{RELATED WORK}

\par The Android platform has become the main target of choice for attackers, posing a serious threat to users' safety and privacy. 
Therefore, it is of great significance to establish a healthy mobile app market.
In recent years, academia and industry have done a lot of important studies for completing the purpose.

 %\textbf{\emph{1) Syntax-based Android Malware Detection}}
Syntax-based methods \cite{peng2012using, aafer2013droidapiminer, wang2014exploring,liu2018large, liu2018mining, arp2014drebin, sarma2012android, zhou2012hey} ignore the semantics of app code to achieve efficient Android malware detection.
For example, Wang \emph{et al.} \cite{wang2014exploring} detects Android malware by analyzing risky permissions requested by an app. 
% It scans the manifest file to collect the list of all permissions, and then applies several feature ranking methods to rank them with respect to the risk. 
% After obtaining the ranking of all analyzed permissions, permissions with top risks will be considered as risky permissions and are used as features to detect malware.
These risky permissions can provide a mechanism of access control to core facilities of the mobile system, so they can be represented as a type of apps' behavior.
Because of the lack of program semantics, such approach suffers from low effectiveness on detecting Android malware.
\emph{Drebin} \cite{arp2014drebin} considers extracting features from both manifest and app code.
%considers extracting eight different types of feature sets from an app, four of which are extracted from the manifest file, and the others are obtained from app code. Feature sets in the manifest consist of hardware features, requested permissions, app components, and filtered intents while includes restricted API calls, used permissions, suspicious API calls, and network addresses in app disassembled code. 
After extracting all the features, it embeds them into a joint vector space to train an SVM to detect malware.
However, it only searches for the presence of particular strings, such as some restricted API calls, rather than considers the program semantics. So it can be easily evaded by attacks on syntax features \cite{chen2018android}.

% \textbf{\emph{2) Semantics-based Android Malware Detection}}
In order to maintain high effectiveness on detecting Android malware, researchers \cite{zhang2014semantics, feng2014apposcopy, mariconti2016mamadroid, yang2015appcontext, allen2018improving,yang2017malware, yang2018enmobile, avdiienko2015mining, machiry2018using, repack1, repack2, repack3, repack4, repack5, repack6} conduct program analysis to extract different types of app semantics.
For example, \emph{MaMaDroid} \cite{mariconti2016mamadroid} first performs static analysis to obtain the graph representation of an app, then all the sequences are obtained from the graph and are abstracted into the corresponding packages to model the app's invocation behaviors. 
% It builds a Markov chain model to represent the transition probabilities between functions. 
% Markov chains stand for multiple pairs of call relationships performed by an app and are used to construct feature vectors to detect malware.
In practice, the abstraction of \emph{MaMaDroid} may bring some false alarms, for instance, \emph{android.telephony.TelephonyManager.getDeviceId()} and \emph{android.telephony.SmsManager.sendTextMessage()} are both abstracted in \emph{android.telephony} package while their usages and corresponding security-levels are completely different.
The main ideas of \emph{DroidSIFT} \cite{zhang2014semantics} and \emph{Apposcopy} \cite{feng2014apposcopy} are similar, that is, they both first extract graph representation of an app, and then conduct graph matching to detect malware.
However, since heavy-weight program analyses are both performed by \emph{DroidSIFT} and \emph{Apposcopy} to obtain accurate graphs, they all suffer from low scalability.
In their papers \cite{zhang2014semantics, feng2014apposcopy}, they report that the average runtime overhead are 175.8 seconds and 275 seconds for \emph{DroidSIFT} and \emph{Apposcopy}, respectively.

% \textbf{\emph{3) Repackaged Android Malware Detection}} 
% Some approaches \cite{repack1, repack2, repack3, repack4, repack5} aim to detect repackaged Android malware, for example, \emph{MassVet} \cite{repack4} builds a viewgraph to describe an app with a reasonably complicated UI structure. 
% To ensure the high scalability on graph matching, \emph{MassVet} applies a similarity comparison algorithm that appeared in their former work \cite{chen2014achieving} to the analysis of the recovered view graph. 
% It has validated the high efficiency and scalability of mobile malware detection, however, the original purpose of \emph{MassVet} is to detect repackaged malware. 
% It can cause a false negative when the app is a new malware.
% \emph{DR-Droid} \cite{repack2} divides the entire app code into multiple dependency-based regions to detect repackaged malware.
% The way they partition the app is to generate a class-level dependency graph by exploring different categories of dependencies.
% After obtaining the partitioned regions, \emph{DR-Droid} extracts features from each region independently and label it as either benign or malicious.
% As long as any region is detected as malicious, the entire app will be considered as a repackaged malware.
% The premise of \emph{DR-Droid} to conduct app partition is that the connections between the normal part and the malicious part are weak.
% Therefore, when detecting covert malware, it may cause false negatives since the normal part and malicious part are highly correlated.

\section{CONCLUSION}

\par In this paper, we generate the first dataset of Android covert malware. 
To detect these covert malware samples, we design a new technique to discover the most suspicious part of covert malware by analyzing the homophily of a call graph.
We implement a prototype system, \emph{HomDroid}, a novel and automatic system that can accurately detect Android covert malware.
We conduct evaluations using 4,840 benign samples and 3,358 covert malicious samples. 
Experimental results show that \emph{HomDroid} is capable of detecting Android covert malware with a False Negative Rate of 3.2\% while are 30.7\%, 16.3\%, 15.2\%, and 10.4\% for four comparative systems (\ie  \emph{PerDroid} \cite{wang2014exploring}, \emph{Drebin} \cite{arp2014drebin}, \emph{MaMaDroid} \cite{mariconti2016mamadroid}, and \emph{IntDroid}~\cite{zou2021intdroid}).

\section*{ACKNOWLEDGEMENTS}

We would thank the anonymous reviewers for their insightful comments to improve the quality of the paper. 
This work is supported by the Key Program of National Science Foundation of China under Grant No. U1936211, `the Fundamental Research Funds for the Central Universities', HUST: 2020JYCXJJ068, and UT Dallas startup funding \#37030034.
%by the Shenzhen Fundamental Research Program under Grant No. JCYJ20170413114215614 and the Key-Area Research and Development Program of Guangdong Province under Grant No. 2019B010139001.

\bibliographystyle{ACM-Reference-Format}

\balance
\bibliography{HomDroid}

\end{document}